\documentclass[usenatbib]{mn2e}
\usepackage{epsfig}
\usepackage{subfig, amssymb}

\def\msun{{\rm M_{\odot}}}

\def\me{{\dot M_{\rm Edd}}}

\def\mo{{\dot M_{\rm out}}}
\def\le{{L_{\rm Edd}}}

\newcommand{\be}{\begin{equation}}
\newcommand{\ee}{\end{equation}}

\def\ltsima{$\; \buildrel < \over \sim \;$}
\def\simlt{\lower.5ex\hbox{\ltsima}}
\def\gtsima{$\; \buildrel > \over \sim \;$}
\def\simgt{\lower.5ex\hbox{\gtsima}}
\def\sgra{Sgr~A$^*$}

\newcommand\ledd{{L}_{\rm Edd}}

\newcommand\mbh{{\,{\rm M}_{\rm bh}}}

\newcommand{\apj}{ApJ}
\newcommand{\apjs}{ApJS}
\newcommand{\mnras}{MNRAS}
\newcommand{\aap}{A\&A}
\newcommand{\araa}{ARA\&A}
\newcommand{\apjl}{ApJL}

\newcommand{\nat}{Nature}

\newcommand{\physrep}{Phys.Rep.}
\def\del#1{{}}

\title[{\it Fermi} Bubbles: \sgra\ AGN feedback?] {{\it Fermi} Bubbles in the
  Milky Way: the closest AGN feedback laboratory courtesy of \sgra?}

\author[Kastytis Zubovas, Sergei Nayakshin] {\parbox{18cm}{Kastytis
    Zubovas$^{*}$ and Sergei Nayakshin}\vspace{0.3cm}\\
\noindent Dept. of Physics \& Astronomy, University of Leicester, Leicester, LE1 7RH, UK}

\begin{document}

\maketitle

\begin{abstract}
Deposition of a massive ($10^4$ to $10^5 \msun$) giant molecular cloud (GMC)
into the inner parsec of the Galaxy is widely believed to explain the origin
of over a hundred unusually massive young stars born there $\sim 6$~Myr ago.
An unknown fraction of that gas could have been accreted by \sgra, the
supermassive black hole (SMBH) of the Milky Way. It has been recently
suggested that two observed $\gamma$-ray-emitting bubbles emanating from the
very center of our Galaxy were inflated by this putative activity of \sgra. We
run a suite of numerical simulations to test whether the observed morphology
of the bubbles could be due to the collimation of a wide angle outflow from
\sgra\ by the disc-like Central Molecular Zone (CMZ), a well known massive
repository of molecular gas in the central $\sim 200$~pc. We find that an
Eddington-limited outburst of \sgra\ lasting $\simeq 1$~Myr is required to
reproduce the morphology of the {\it Fermi} bubbles, suggesting that the GMC
mass was $\sim 10^5 \msun$ and it was mainly accreted by \sgra\ rather than
used to make stars. We also find that the outflow from \sgra\ enforces strong
angular momentum mixing in the CMZ disc, robustly sculpting it into a much
narrower structure -- a ring -- perhaps synonymous with the recently reported
``{\it Herschel} ring''. In addition, we find that \sgra\ outflow is likely to
have induced formation of massive star-forming GMCs in the CMZ. In this
scenario, the Arches and Quintuplet clusters, the two observed young star
clusters in the central tens of parsecs of the Galaxy, and also GMCs such as
Sgr B2, owe their existence to the recent \sgra\ activity.
\end{abstract}

\renewcommand{\thefootnote}{\fnsymbol{footnote}}
\footnotetext[1]{E-mail: {\tt kastytis.zubovas@astro.le.ac.uk }}

\section{Introduction}

The Galactic centre (GC) is a particularly interesting astrophysical location,
primarily due to the presence of the closest supermassive black hole (SMBH) -
\sgra, whose mass is $\mbh \simeq 4 \times 10^6 \msun$
\citep{Schodel2002Natur, Ghez2005ApJ, Ghez2008ApJ}. Although the Soltan
relation \citep{Soltan1982MNRAS} implies that \sgra\ gained most of its mass
through luminous accretion, it is currently very dim in comparison with active
galactic nuclei (AGN): its X-ray luminosity is less than $\sim 10^{-11} \ledd$
(where $\ledd \sim$ a few $\times 10^{44}$~erg~s$^{-1}$ is its Eddington
luminosity; \citealt{Baganoff2003ApJ}). \sgra\ dimness has stimulated
development of radiatively inneficient accretion and accretion/outflow
solutions \citep[e.g.,][]{NarayanEtal95,BB99}. However, X-ray reflection
nebulae suggest that \sgra\ might have been much brighter about a hundred
years ago, with luminosity of a few $\times 10^{39}$~erg~s$^{-1}$  in
  X-rays \citep[e.g.][]{Revnivtsev2004A&A, Ponti2010ApJ}. This variability may
reflect feeding events from a few pc-scale molecular gas reservoirs
\citep{Morris1999AdSpR} or variability in the wind capture rate from the young
massive stars near \sgra\ that feed it presently \citep{Cuadra2008MNRAS}.

A recent observation may shed a rather unexpected light on a much earlier but
still rather recent activity of \sgra. {\it Fermi}-LAT data has been recently
analysed to reveal two giant $\gamma$-ray emitting bubbles, disposed
symmetrically on either side of the Galactic plane (\citealt{Su2010ApJ};
although see \citealt{Dobler2011ApJ} for a different interpretation). They are
roughly teardrop-shaped and extend $\sim 8-10$~kpc above and below the plane,
but are centred on \sgra\ with a narrow ($d \sim 100$~pc) waist along the
plane. The limbs of the lobes coincide with the extended structure seen in
medium-energy X-rays by ROSAT \citep{Snowden1997ApJ}.  The lobes have
$\gamma$-ray luminosity $L_{\gamma} \simeq 4\times 10^{37}~{\rm erg\,
  s}^{-1}$. Observational constraints \citep{Su2010ApJ} and emission modelling
\citep{Crocker2011PhRvL} allows one to estimate the total kinetic and thermal
energy content of the {\it Fermi} bubbles as $E_{\rm bub} \sim 10^{54-55}$~erg.

There has been a number of suggestions as to what the origin of the {\it
  Fermi} bubbles might be. \citet{Su2010ApJ} discuss several physical
processes and provide a constraint that if the bubbles are older than a few
$\times 10^6$~yr, the $\gamma$-ray emission must be powered by ions rather
than electrons due to a short cooling time of the latter (see their section
7.1 and Figure 28), unless electrons are continuously accelerated within
  the bubbles (which may not be unreasonable; see \S \ref{sec:radiation}
  below). \citet{Crocker2011MNRAS} and \citet{Crocker2011PhRvL} detailed
these arguments further and suggested that the emission is powered by cosmic
ray (CR) protons rather than electrons. They further consider a quasi-steady
state model in which the CR protons are continuously injected by supernova
explosions. CR protons and heavier ions are then trapped inside the bubbles
for the lifetime of the latter (which the authors require to be about 10
Gyrs). On the other hand, \citet{Mertsch2011PhRvL} argued that the emission
spectrum of the bubbles is inconsistent with CR protons and therefore
the bubbles must be a recent, cosmic ray electron-powered, feature.

The scenario with a more recent origin of the bubbles has been investigated by
several authors. \citet{Guo2011arXiv} suggest that a jet launched by
\sgra\ $1-2$~Myr ago could create the morphology and emission structure
observed. \citet{Cheng2011ApJ} argue that the bubbles are inflated by episodic
\sgra\ activity caused by tidal disruptions of stars passing too close to
\sgra. \citet[hereafter Paper I]{Zubovas2011MNRAS} argue that an almost
spherical outflow from \sgra\, caused by a short burst of AGN activity
coincident with the well-known star formation (SF) event $6$~Myr ago
\citep{Paumard2006ApJ}, can be the origin of the bubbles. In the model, the
outflow from \sgra\ is suggested to be quasi-spherical at launch as is
required to explain the observed $\mbh-\sigma$ relations for classical bulges
and elliptical galaxies \citep{King2003ApJ, King2005ApJ, King2010MNRASa}. The
outflow becomes collimated by interaction with a disc-like structure, the
Central Molecular Zone - a dense ring (aspect ratio $H/R \sim 0.2-0.3$) of
predominantly molecular gas extending from the Galaxy centre to
$\sim200$~pc. This collimation explains the large width of the bubbles and
their symmetry with respect to the Galactic plane, which is not at all
guaranteed by a jet model, since jet directions are known to be completely
uncorrelated with the large-scale structure of the host galaxies
\citep{Kinney2000ApJ, Nagar1999ApJ}. \cite{Zubovas2011MNRAS} estimated the
duration of the quasar phase to be at the minimum $t_q = 5 \times 10^4$~yr and
the mass of gas accreted by \sgra, $\Delta M > 4 \times 10^3 \msun$,
comparable to the mass in the young stars in the central parsec, $M_* \sim $
few to ten $\times 10^3 \; \msun$ \citep{Paumard2006ApJ}.

In this paper, we test the analytical results of Paper I by performing
numerical simulations. We broadly follow the methodology of simulating
spherical outflows from accreting SMBHs of \citet{Nayakshin2010MNRAS}. These
SPH simulations involve a prescription for passing momentum and energy of the
outflow to the ambient gas \citep{Nayakshin2009MNRASb}. The method has been
shown \citep{Nayakshin2010MNRAS} to reproduce the analytical results of the
wind feedback model we use here \citep{King2003ApJ}. We aim to investigate
whether the morphology, size and energy content of the bubbles can be
reproduced with this model. In addition, we are interested in constraints that
the simulations could place on a key unknown of the problem -- the duration of
\sgra\ activity. We find that a longer quasar outburst is needed than found in
Paper I, most likely because the energy produced by \sgra\ outflow is spread
more widely in the simulations, e.g., outside the bubbles themselves, than
assumed analytically.

The Paper is structured as follows. In Section \ref{sec:model}, we review our
analytical model and the  main results of Paper I. In Section \ref{sec:numsim},
we describe the numerical model and the initial conditions of the
simulations. We follow this by presenting the results in Section
\ref{sec:results} and discussing their implications in Section
\ref{sec:discuss}. We briefly summarize and conclude in Section
\ref{sec:concl}.

\section{A physical model for the {\it Fermi} bubbles} \label{sec:model}

\subsection{\sgra\ as the power source}\label{sec:power}

As explained in Paper I, our \sgra\ feedback model is based on the analytical
models of fast wind outflows from supermassive black holes by
\citet{King2003ApJ, King2005ApJ, King2010MNRASb, King2010MNRASa}, tailored to
the present day inner Milky Way.  Here we summarise the key assumptions behind
this model.

Fast ($v\sim 0.1 c$) wide--angle outflows from AGN have been revealed by
observations of blue--shifted X-ray absorption lines; they appear to be a
ubiquitous feature, detected in $\sim40\%$ of AGN \citep{Tombesi2010A&A,
  Tombesi2010ApJ}. Recent XMM observations \citep{PoundsVaughan11} of X-ray
emission from a limb-brightened shell of post-shock gas building up ahead of
the contact discontinuity corroborate these results and the model of
\citet{King2003ApJ} further. The outflows are probably driven by radiation
pressure of photons emitted by \sgra\ on the gas in the innermost regions of
black hole accretion flows. While some anisotropy in the outflow may be
present in general, here we consider a spherical outflow for simplicity and
demonstrate that it is the properties of the host galaxy that may mold the
outflows into the teardrop-like shape.

We assume that \sgra\ accretion rate is mildly super-Eddington, and that its
luminosity is limited by the Eddington limit. We do not model the sub-parsec
scale disc(s) from which \sgra\ would be accreting gas here. As shown by
\cite{Nayakshin2005A&A} and \cite{AlexanderEtal12a}, viscous disc time scales
are long enough to maintain \sgra\ feeding via an accretion disc for at least
a million years. Furthermore, Nayakshin et al (2012, ApJ submitted) show that
AGN feedback is unable to remove gas discs {\em within} the black hole's
sphere of influence. Physically, this result is easily understood: escape
velocities for gas in the sub-parsec scale disc are much larger than
that for gas in the bulge of the host, so much more energy needs to be
expended to eject gas from very near the black hole than from the bulge of the
Galaxy.  This is why a relatively low mass sub-resolution reservoir of gas (an
accretion disc) could power \sgra\ for the duration of the quasar outburst
envisioned here, while in the meantime \sgra\ drives much more massive
outflows on scales of the Galaxy's bulge.

The outflow is believed to be self-regulating to have a scattering optical
depth $\sim 1$ \citep{King2010MNRASa}, so that the photons transfer their
momentum to the outflowing wind which carries momentum
\begin{equation}
\mo v \simeq \frac{\le}{c},
\end{equation}
where $v \sim \eta c$, and $\eta \simeq 0.1$ is the accretion efficiency.  We
note that the mass outflow rate from \sgra, $\mo$, is bound to be of the order
of $\sim \me$, which is negligible on the Galactic scales, implying that we
can neglect injection of the mass into the Galaxy (but not the outflow's
momentum or the energy, as we discuss now).

\subsection{The effects of the outflow on the diffuse gas in the
  Galaxy}\label{sec:king03} 

Since the outflow velocity is very much larger than the escape velocity from
the Galaxy, the outflow propagates with an approximately constant velocity
until it shocks against the ambient gas. For a sufficiently powerful outflow,
the shocked wind continues to drive an outflow outward, clearing the inner
regions of the bulge of gas.  The $M_{\rm BH} - \sigma$ relation for
elliptical galaxies and classical bulges may have been established by such
outflows that terminated both host galaxy and SMBH growth \citep{King2005ApJ}.

The point that we want to emphasize now is that even SMBHs lying well below
this relation can drive the gas out of their host bulges, provided that the
gas density is lower than that expected in the gas rich era when the bulge
itself forms \citep{King2003ApJ,King2005ApJ}. In our model, gas density in the
bulge is parametrised through the gas fraction $f_{\rm g}$: \be\label{eq:rho}
\rho_{\rm g} = \frac{f_{\rm g} \sigma^2}{2 \pi G R^2} = 2.5 \times 10^{-26}
f_{-3} R_{\rm kpc}^{-2} \; {\rm g}\;{\rm cm}^{-3}, \ee where $\sigma$ is the
velocity dispersion in the bulge and $R$ is the radial coordinate. The
$\propto R^{-2}$ scaling of the density is natural as this is the same as that
for the background stellar density that dominates the potential in the bulge.
We further parametrize $f_{\rm g} \equiv 10^{-3}f_{-3}$, $R = R_{\rm kpc}$~kpc
and we use the value $\sigma = 100$~km/s, appropriate for the Milky Way. For
reference, the model of \cite{King2003ApJ,King2005ApJ} assumes that $f_g =
0.16$, the cosmological value, during the bulge formation epoch.

The dynamics of the outflow depends crucially on whether the shocked wind cools
efficiently or not. The shock temperature is of the order of $10^{11}$~K, so
the only effective form of cooling is the inverse Compton effect of the
shocked outflow's electrons on the \sgra\ radiation field \citep[which would
  dominate over the whole Galaxy's light;][]{Ciotti1997ApJ}. Outside a certain
radius, $R_{\rm cool}$, the radiation field is unable to cool the shock and
the outflow transitions from momentum-driven to energy-driven
\citep{Zubovas2012ApJ}. This radius can be estimated by comparing the cooling
and the flow timescales \citep{King2003ApJ}. For the parameters of the Milky
Way ($M_{\rm BH} = 4\times 10^6 \msun$), $R_{\rm cool}$  is
\begin{equation}
\label{rcool} R_{\rm cool} \simeq 12 \; f_{-3}^{1/2} \; \mathrm{pc}\;.
\end{equation}
Even for the cosmological gas fraction ($f_{-3} = 160$), $R_{\rm cool}$ is
very small compared with the size of the bulge. For the problem at hand, where
the gas density in the bulge is likely to be much smaller than this, this
implies that for all the relevant parameter space any outflow from
\sgra\ quickly becomes energy-driven. In this type of outflow, almost all of
the kinetic power of the outflow,
\begin{equation} \label{eq:dote}
\dot E_{\rm out} = {1\over 2}\mo v^2 = {\eta^2c^2\over 2}\mo = {\eta\over 2}\le,
\end{equation}
is passed to the ambient gas, heating and pushing it outward. The shocked shell
rapidly (in $t \simlt 10^5$~yr) attains a constant velocity, which for a
spherically symmetric outflow is given by
\begin{equation} 
\label{eq:ve}
v_{\rm e} = \left[\frac{2 \eta \sigma^2 c}{3 }\frac{0.16}{f_{\rm
      g}}\frac{M_{\rm BH}}{M_{\sigma}} l \right]^{1/3} \simeq 1920 \;
\sigma_{100}^{2/3} f_{-3}^{-1/3} l^{1/3} \; \hbox{km s}^{-1}\;,
\end{equation}
where $M_{\sigma}$ is the value of the SMBH mass expected from the $M -
\sigma$ relation \citep{King2011MNRAS}, $M_{\rm BH} \simeq 0.2 M_\sigma$ and
$l \sim1$ is the Eddington ratio.

Once the quasar switches off, the shell continues to expand, but stalls and
eventually stops (assuming that the isothermal potential of the bulge
extends to infinity). The maximum radius which the shell reaches is
\begin{equation}
R_{\rm stall} \simeq {v_{\rm e}\over \sigma}R_0 \simeq {v_{\rm e}^2\over
  \sigma}t_{\rm q},
\label{rstall}
\end{equation}
where $t_{\rm q}$ is the duration of the quasar outburst. 

Using the size of the observed {\it Fermi} bubbles ($R_{\rm stall} \simgt
10$~kpc) we can constrain the free parameters of our model, $f_{-3}$ and
$t_{\rm q}$. We estimate the gas fraction from equation (\ref{eq:ve}),
requiring that the bubbles must have reached the present radius in 6~Myr. This
is equivalent to assuming that the shell's mean velocity is $\left<v\right>
\gtrsim 1600$~km/s (see Paper I for details), and yields
\begin{equation} 
f_{-3} \lesssim \left(\frac{1920}{1600}\right)^{3} l \simeq 2l\;.
\end{equation} 
We see that the gas fraction in the Milky Way halo must have been similar to
$10^{-3}$ in order for the bubbles to be inflated within our model. Numerical
simulations below confirm this.

Requiring that the bubble stalling radius should be greater than their
  current radius, equations (\ref{rstall}) and (\ref{eq:ve}) give
\begin{equation}
 \label{eq:tqanal}
 t_{\rm q} > 2.5 \times 10^5 f_{-3}^{2/3} \; {\rm yr}.
\end{equation}
As the bubbles are likely to be still expanding at the present time (see Paper
I), this result is a lower limit.  We note that the estimate given by equation
\ref{eq:tqanal} is a factor of 5 larger than the estimate obtained in Paper
I. As both estimates are lower limits on the outburst duration, only the
larger one of the two is relevant.

In $0.25$~Myr, \sgra\ accreting at its Eddington limit consumes $\Delta M
\simeq 2\times10^4\msun$ of gas. Within our feedback model, $\sim 2 \times
10^{56}$~erg is ejected in the outflow's mechanical energy. This is more than
an order of magnitude larger than the bubbles energy content estimated by
\citet{Su2010ApJ}.  However, we find that a large fraction of the outflow's
energy goes into mechanical work expended to drive the ambient cooler medium
away from the Galaxy's centre. \citet{Zubovas2012ApJ} show that even while the
quasar driving is on, only $1/3$ of the energy input is retained in the
shocked wind while the quasar is active. When the quasar turns off, the bubble
expands adiabatically, converting its thermal energy into kinetic energy of
the bubble and the surrounding shell. Therefore it is likely that the actual
amount of energy retained by the bubbles is much lower than the original
energy input by \sgra\ into the outflow, bringing the value in line with the
observations. Our simulations confirm this prediction; cf. \S
\ref{sec:energy}.

\subsection{The role of the Central Molecular Zone in focusing the
  outflow} \label{sec:waist} 

The shell expansion velocity, $v_{\rm e}$, depends on the gas density in the
direction of expansion. Clearly, if the ambient gas distribution is not
spherically symmetric then the outflow must lose its spherical symmetry
too. As a minimum effect, the velocity of the contact discontinuity must be
smaller in the directions of denser gas. The most salient feature in the
distribution of gas in the inner Galaxy is that a good fraction of it --
mainly the cold molecular gas -- lies close to the plane of the Galaxy. For
what follows it is the Central Molecular Zone \citep[CMZ;][]{Morris1996ARA&A},
a massive disc-like molecular gas feature in the inner $\sim 200$~pc, that
matters the most. As we argued in Paper I, the CMZ presents an almost impassable
barrier to the \sgra\ outflow. This can be seen from the fact that the weight
of the CMZ,
\begin{equation}
W_{\rm CMZ} \sim {G M_{\rm enc} M_{\rm CMZ} \over R_{\rm CMZ}^2}
= {2 M_{\rm CMZ} \sigma^2 \over R_{\rm CMZ}} \sim 6.5\times 10^{34} \; {\rm
  dyn},
\label{wcmz}
\end{equation}
where $R_{\rm CMZ}$ is the mean radius of the CMZ and $M_{\rm enc}$ is the mass
enclosed within this radius, is greater than the momentum flux impacting on
the CMZ from \sgra\ outflow (which is the outward force on the CMZ):
\begin{equation}
F_{\rm CMZ} \sim \frac{L_{\rm Edd}}{c}\frac{H_{\rm CMZ}}{R_{\rm CMZ}} \sim
2\times 10^{33} \; {\rm dyn}.
\end{equation}
Therefore we expect the CMZ to be a very effective obstacle to the outflow
(this estimate does not take into account energy deposition by the outflow
into the CMZ, which makes the latter somewhat more prone to the feedback; see
\S \ref{sec:cmzshape}). Physically, we expect \sgra\ outflow impacting the CMZ
to shock and thermalise. As the CMZ weight is so large, the outflow is either
completely stopped or nearly stalled in the plane of the Galaxy. The shocked
outflow gas cannot simply pile up there, however. Indeed, if that were the
case then the pressure (thermal energy density) would increase in that
location without limit. The thermal pressure of the shocked gas in the
directions perpendicular to the Galactic plane is much lower because the
outflow proceeds in those directions easily. Thus there is a strong pressure
gradient in the shocked outflow gas pointing towards the Galactic plane. This
pressure gradient clearly must launch a ``secondary'' thermally driven outflow
away from the Galactic plane, efficiently collimating \sgra\ outflow into
these directions.

\begin{table*}
\begin{center}
  \begin{tabular}{c c c c c c || c c c c c c}
    \hline
    \hline
    Test & $f_{\rm g}$ & $N_{\rm p, halo}$ & $t_{\rm q}/$Myr & $\left(\frac{H}{R}\right)_{\rm cmz}$ & $M_{\rm cmz}/\msun$ & $N_{\rm p,cmz}$ & $R_{\rm b} /$kpc & $d_{\rm b} /$kpc & $h_{\rm b} /$kpc & $v_{\rm v} /$km s$^{-1}$ & $v_{\rm h} /$km s$^{-1}$\\
    \hline
    Base    & $10^{-3}$         & $7.5 \times 10^5$ & $1$  & $0.25$ & $10^8$ & $10^6$ & $11.5$ & $9$  & $\sim1$ & $1090$ & $670$ \\

    HR-low  & $10^{-3}$         & $7.5 \times 10^5$ & $1$  & $0.125$& $10^8$ & $10^6$ & $11.5$ & $9$ & $\sim1$ & $1110$ & $680$ \\

    Cool$^*$& $10^{-3}$         & $7.5 \times 10^5$ & $1$  & $0.25$ & $10^8$ & $10^6$ & $9.5$  & $10$ & $\sim1$ & $820$ & $650$ \\

\hline

    Fg-low  & $4 \times 10^{-4}$ & $3.0 \times 10^5$ & $1$  & $0.25$ & $10^8$ & $10^6$ & $\sim15$  & $\sim10$ & $2.5$ & $2560$ & $610$ \\

    Fg-high & $4 \times 10^{-3}$ & $3.0 \times 10^6$ & $1$  & $0.25$ & $10^8$ & $10^6$ & $6$   & $6$ & $\sim1$ & $220$ & $400$ \\

    Tq-low  & $10^{-3}$         & $7.5 \times 10^5$ & $0.3$& $0.25$ & $10^8$ & $10^6$ & $7$    & $4$  & $2$ & $300$ & $360$ \\

    Both-low& $4 \times 10^{-4}$ & $3.0 \times 10^5$ & $0.3$& $0.25$ & $10^8$ & $10^6$ & $\sim7$    & $\sim4$  & $4$ & $960$ & $270$ \\

    Mc-low  & $10^{-3}$         & $7.5 \times 10^5$ & $1$  & $0.25$ & $10^7$ & $10^5$ & $7$    & $8$  & $2$ & $330$ & $470$ \\

    \hline
    \hline
  \end{tabular}
  \caption[Parameters and results]{Simulation parameters and main
    results. From left to right, the parameters are: Test ID, gas fraction,
    number of SPH particles in the halo, quasar outburst duration, CMZ mass,
    CMZ scale height, number of particles in the CMZ. The results are: bubble
    height, width and distance between its lower edge and the SMBH; velocity
    of the swept-up ISM in the $z$ direction at $x = y = 0$ and velocity of
    the swept-up ISM in the $xy$ plane at the mid-height of the bubble; all five
    at $6$~Myr.

    $^*$ - Simulation `Cool' is identical to `Base', but includes a
    \cite{Sazonov2005MNRAS} heating-cooling prescription. }

  \label{table:paramsnew}
\end{center}
\end{table*}

\section{Simulation setup} \label{sec:numsim}

\subsection{Numerical method} \label{sec:scheme}

Our workhorse code for solving gas dynamics in the fixed potential of the bulge and
the black hole is GADGET-3, an updated version of the code presented in
\citet{Springel2005MNRAS}. Feedback from the SMBH is implemented with the
'virtual particle' method explained in detail in
\citet{Nayakshin2009MNRASb}. Rather than hydrodynamically modelling the
low-density fast wind flowing from the SMBH, its effects on the gas
distribution at large are simulated via a distinct population of particles,
called `virtual particles' since they exist only during their propagation from
the source, \sgra, to the point of interaction with the ambient gas. These
particles are ejected isotropically from the SMBH at each of its timesteps and
move radially with a set constant velocity ($v_{\rm w} = 0.1c$). Each particle
carries linear momentum 
\begin{equation}
\label{eq:pvirt} p_\gamma = \frac{L_{\rm Edd}  t_{\rm BH}}{c N_\gamma} 
\end{equation}
and kinetic energy $E_{\gamma} = p_\gamma v_{\rm w}/2$, where $t_{\rm BH}$ is
the SMBH timestep, and divided equally among $N_\gamma$ virtual particles
ejected in that timestep. The scheme is designed to eject the correct momentum
flux from the SMBH, $\ledd/c$, and the number of particles ejected in a time
step, $N_\gamma$, is selected so that the interactions with SPH particles are
sufficiently frequent to reduce the (random-number-generated) noise to acceptable
levels \citep[see][]{Nayakshin2009MNRASb}.

Each virtual particle has a search radius (a multiple of the smoothing length
of an SPH particle), within which search for potential SPH neighbors is
performed ``on the fly''. Whenever an SPH particle kernel contains a virtual
particle, the two are considered interacting, and a fraction of the virtual
particle's momentum and energy is transferred to the gas.  When the virtual
particle energy drops below $1\%$ of its original energy, the particle is
destroyed.

The kinetic energy of the virtual particles is passed to the SPH particles if
the shocked wind does not cool, i.e. if $R > R_{\rm cool}$ (eq. \ref{rcool}).
We have implemented a step function transition between the pure momentum
feedback regime and the energy feedback one for this. However, testing showed
that since $R_{\rm cool}$ is very small in the case considered here, there is
no significant difference in results between simulations utilising the switch
and those employing the energy feedback scheme throughout the simulation, so
we use the latter below.

While we do not model hydrodynamically the reverse wind shock, which
thermalizes the wind and creates a pressurized bubble that can expand
adiabatically, we find that this process is somewhat mimicked in the
simulations by low density gas that is present in the otherwise evacuated
bubble. These particles predominantly originate on the surfaces of the Central
Molecular Zone (see Sections \ref{sec:galmod} and \ref{sec:Base_centre}
below), where they are heated to very high temperatures and rise to fill the
voids; subsequently quasar wind heats these particles even further. The
bubbles are found to be significantly overpressurized with respect to the
surrounding gas. This allows the bubble to expand thermally as expected if the
reverse shock were modelled by SPH (see Section \ref{sec:Base} below).

In all the simulations presented below, we use an ideal equation of state for
the gas. The gas pressure is given by $P= \rho k T/\mu$, where mean molecular
weight, $\mu = 0.63 m_{\rm p}$ (assuming ionised gas of Solar abundance), $k$
is the Boltzmann's constant, and $\rho$ and $T$ are the gas density and
temperature, respectively. In all simulations except for one, we employ the
standard GADGET optically thin cooling prescription based on the
\citet{Sutherland1993ApJS} cooling curves, in addition to heating from the
virtual particles on the contact discontinuity of the bubble. In the only
exception to this, simulation `Cool' (see Table \ref{table:paramsnew}), we
check the sensitivity of our results to the assumed cooling function by
utilising the optically thin radiative cooling rates for gas ionised (and also
heated) by the quasar radiation field as calculated by
\citet{Sazonov2005MNRAS}, who also provided an analytical fit to the
respective rates that we use here.

Simulation snapshots showing density and temperature are plotted using an
angle slice projection method presented in \citet{Nayakshin2010MNRAS}.
Specifically, the gas column density projected over the $y$ coordinate is
calculated by
\be
\Sigma(x,z) = \int_{-y(x,z)}^{y(x,z)} \rho(x,y,z) \, dy,
\ee
where the limits of the integration are given by $y(x,z) = r \, \mathrm{tan}
\, \zeta$ and $r = \sqrt{x^2 + z^2}$. The angle $\zeta$ is chosen so that
$\mathrm{tan} \, \zeta = 1/4$ throughout this paper. This projection method
conveniently allows us to get an unobscured look into the inner
parts of the simulation and yet have enough particles at the outer edges for a
statistically meaningful figure to be derived when plotting. In addition,
since most of the results presented in this paper are symmetrical around the
$z$ axis, the snapshot plots, where applicable, are divided vertically and
show the surface density in blue-white on the left and temperature in
red-orange on the right.

\subsection{Galaxy model and initial conditions} \label{sec:galmod}

The initial setup for the Galaxy consists of three components, described below
and summarized in Table \ref{table:paramsnew}. The `Base' model is the model
which best reproduces the {\it Fermi} bubble observations \citep{Su2010ApJ}. We
analyze this simulation in detail below, and study the robustness of our
results by varying the free parameters of the model.

The whole computational domain is embedded in a static isothermal background
potential with $\sigma = 100$~km/s. In the centre of the coordinate system,
fixed in space, is the SMBH. While we do not model accretion onto \sgra\ in
detail (cf. \S \ref{sec:power}), ocasionally SPH particles may get very close
to the SMBH, especially closer to the end of the simulations when feedback
from \sgra\ is turned off. To avoid very small time steps, and thus very high
numerical costs, associated with these ``uninteresting'' particles, we remove
them using the sink particle formalism if the SPH particles are closer than
$r_{\rm sink} = 0.1$~pc to \sgra. As explained above, \sgra\ is ``turned on''
at the start of each simulation, radiating at its Eddington limit for a
duration $t_{\rm q}$, which is a free parameter of the model. Most of the
models presented here use $t_q = 1$~Myr (since this was found to give the best
results), although we explore shorter outbursts as well.

As mentioned above, we position a massive disc of gas -- the Central Molecular
Zone \citep[CMZ;][]{Morris1996ARA&A} -- in the plane of the Galaxy at a
distance between $R_{\rm in,cmz} = 5$~pc and $R_{\rm out,cmz} = 200$~pc from
\sgra. The inner radius is chosen so that the whole CMZ would be comfortably
outside the sphere of influence of \sgra ($R_{\rm infl} \sim 2$~pc), but still
account for the yet smaller circumnuclear disc of gas, which is observed to
lie between $\sim2$ and $\sim10$ pc from the centre \citep{Guesten1987ApJ,
  Morris1996ARA&A}. Our model CMZ has uniform temperature throughout, chosen
to give the appropriate constant scale height aspect ratio ($H/R = 0.25$ or
$0.125$, with a corresponding difference in the adopted temperature floor for
the simulations). It is rotationally supported in the radial direction, with
$v_{\rm rot} = \sqrt{2} \sigma$. The radial density distribution follows a
$\rho \propto R^{-2}$ power law, and the mass of the disc is set to $M_{\rm
  cmz} = 10^8 \msun$ (approximately the upper limit from observational
constraints), although we ran one simulation with a much lower CMZ
mass. Despite the high mass of the CMZ, it is marginally stable to
self-gravity, since the \cite{Toomre64} parameter $Q \sim 3.4$ for $H/R =
0.25$, and $1.7$ even for $H/R = 0.125$. We note that we do not fine-tune the
CMZ to be only marginally stable in the simulations; it is a natural outcome
of using the observationally constrained CMZ parameters. This result is
probably not a coincidence: massive cold discs are widely believed to be
self-regulating to have $Q \sim1$
\citep{Goodman2003MNRAS,Thompson2005ApJ,Nayakshin2012MNRAS}. In principle,
such a low Q parameter might lead to fragmentation. However, since we set
  the temperature floor in the simulations to be equal to the initial CMZ
  temperature, we do not expect a significant fragmentation to occur
\citep{Gammie01,Rice05}. This assumption is consistent with simulation
  results (see Sections \ref{sec:Base_centre} and \ref{sec:cmzshape} below).
The SPH particle mass of the simulations is $M_{\rm part} = 100 \msun$, giving
a minimum mass resolution of $\sim 4000 \msun$.

\begin{figure*}
  \centering
    \subfloat{\includegraphics[width=0.45\textwidth]{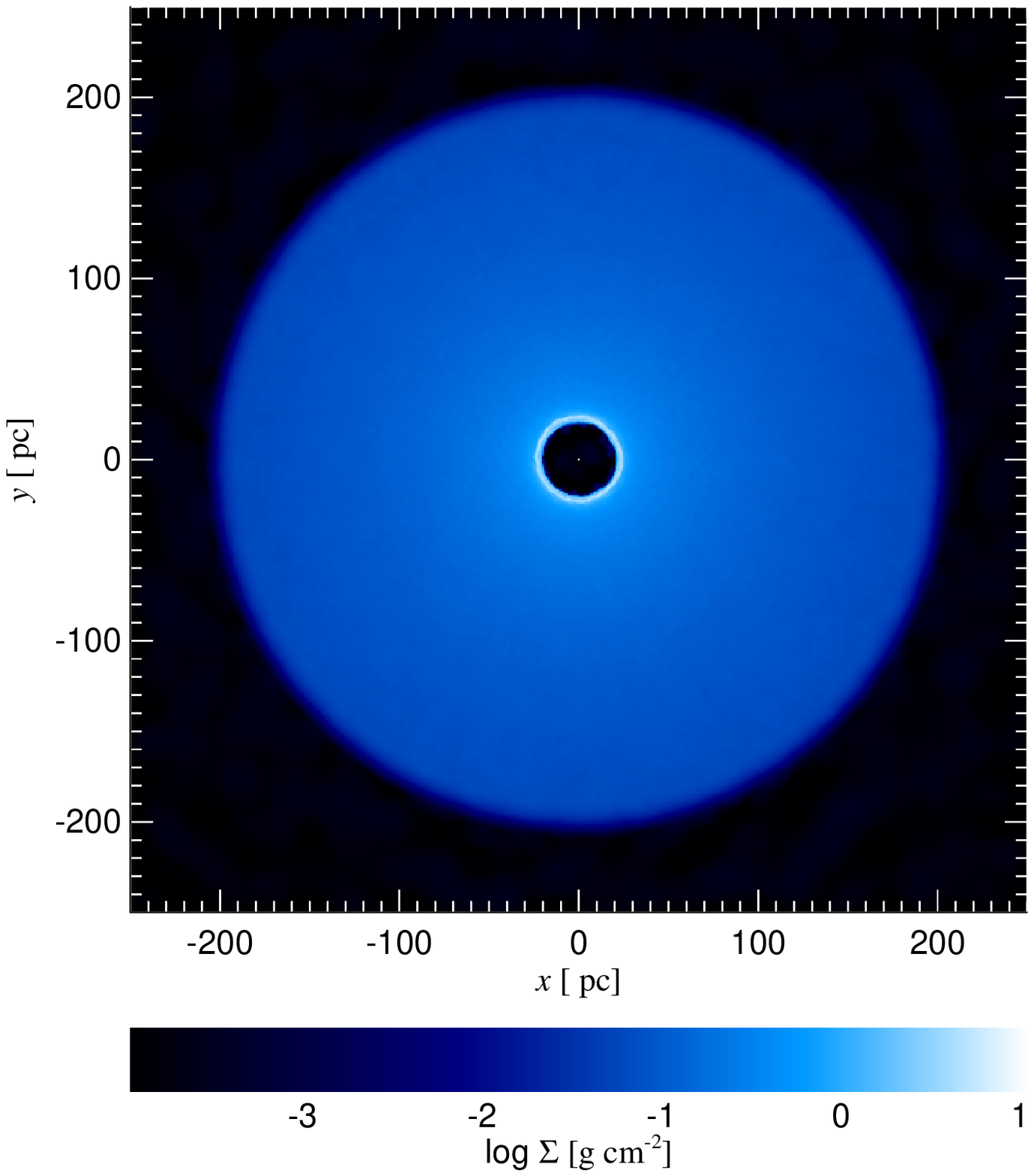}}
    \subfloat{\includegraphics[width=0.45\textwidth]{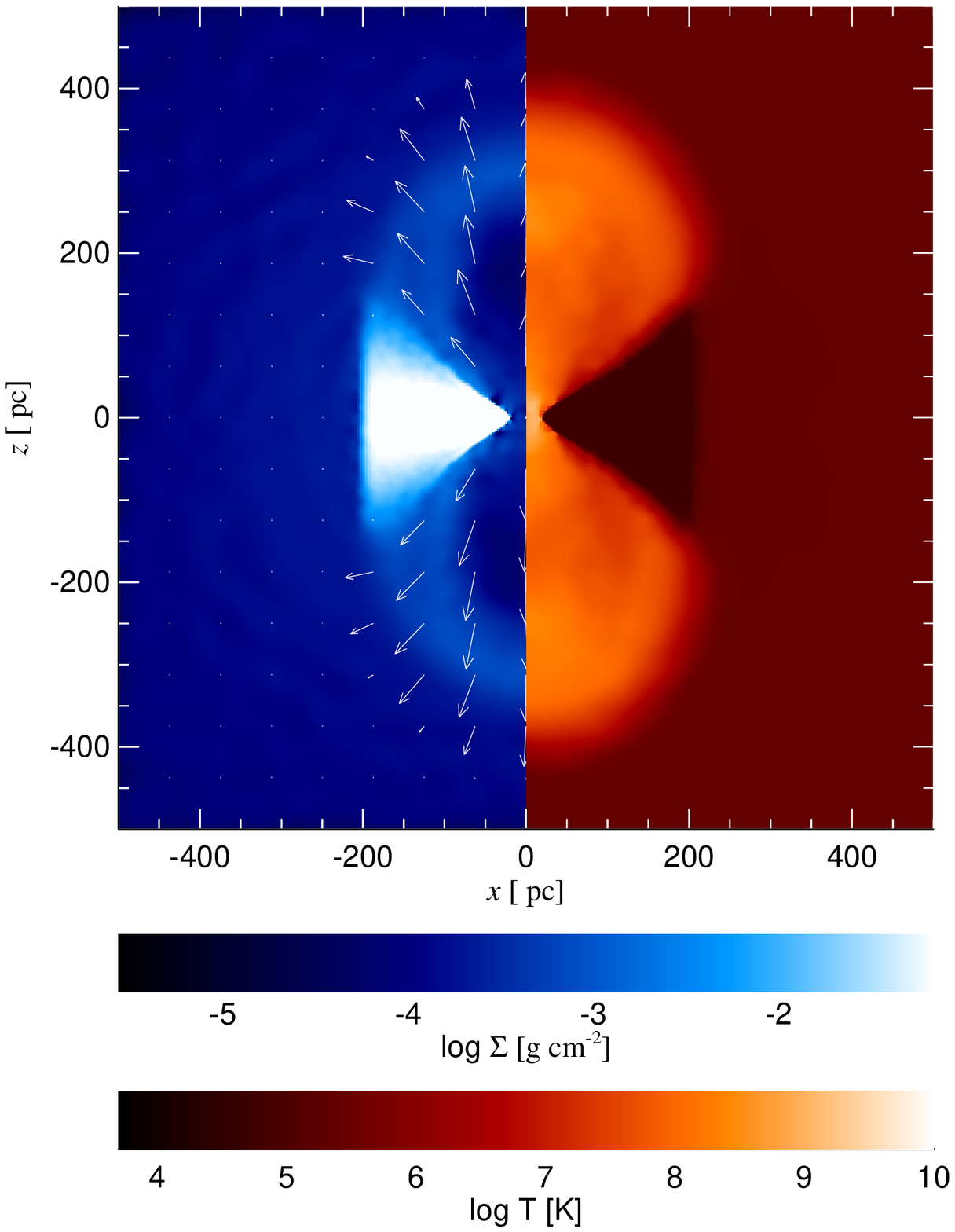}}
  \caption[Early feedback effects in the central regions]{{\bf Left:} The
    face-on column density of the CMZ disc at time $t=0.1$~Myr for the `Base'
    simulation. Note that the innermost region has been partially evacuated by
    \sgra\ feedback. {\bf Right:} Cross-sectional plot of gas surface density
    (left half of the panel, blue-white) and temperature (right, red-orange)
    for the same snapshot. The CMZ (cold dense wedge in the Galactic plane)
    strongly collimates the outflow; even though its surfaces are ablated,
    teardrop-shaped cavities form readily.}
  \label{fig:Base_early1}
\end{figure*}

\begin{figure*}
  \centering
    \subfloat{\includegraphics[width=0.45\textwidth]{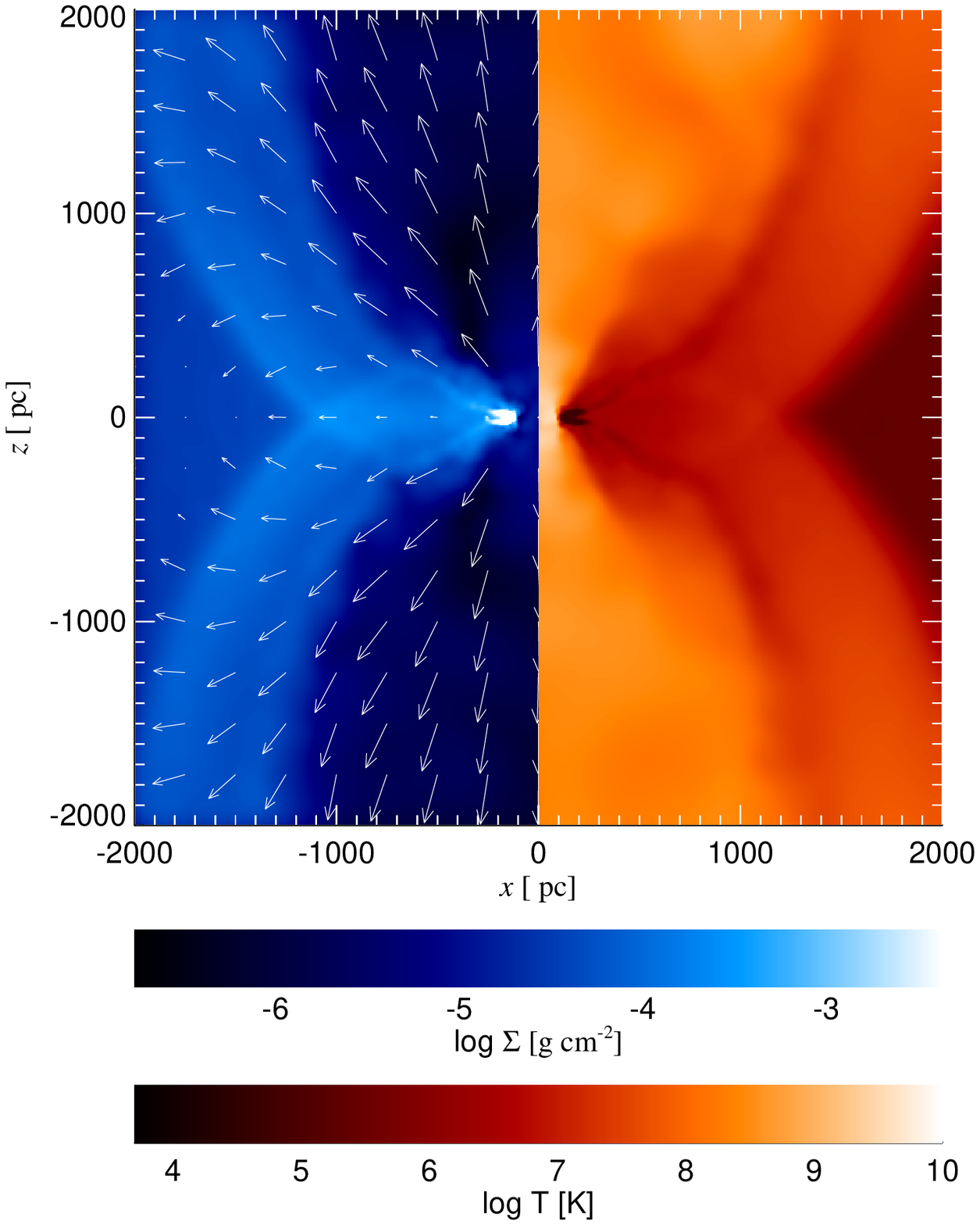}}
    \subfloat{\includegraphics[width=0.45\textwidth]{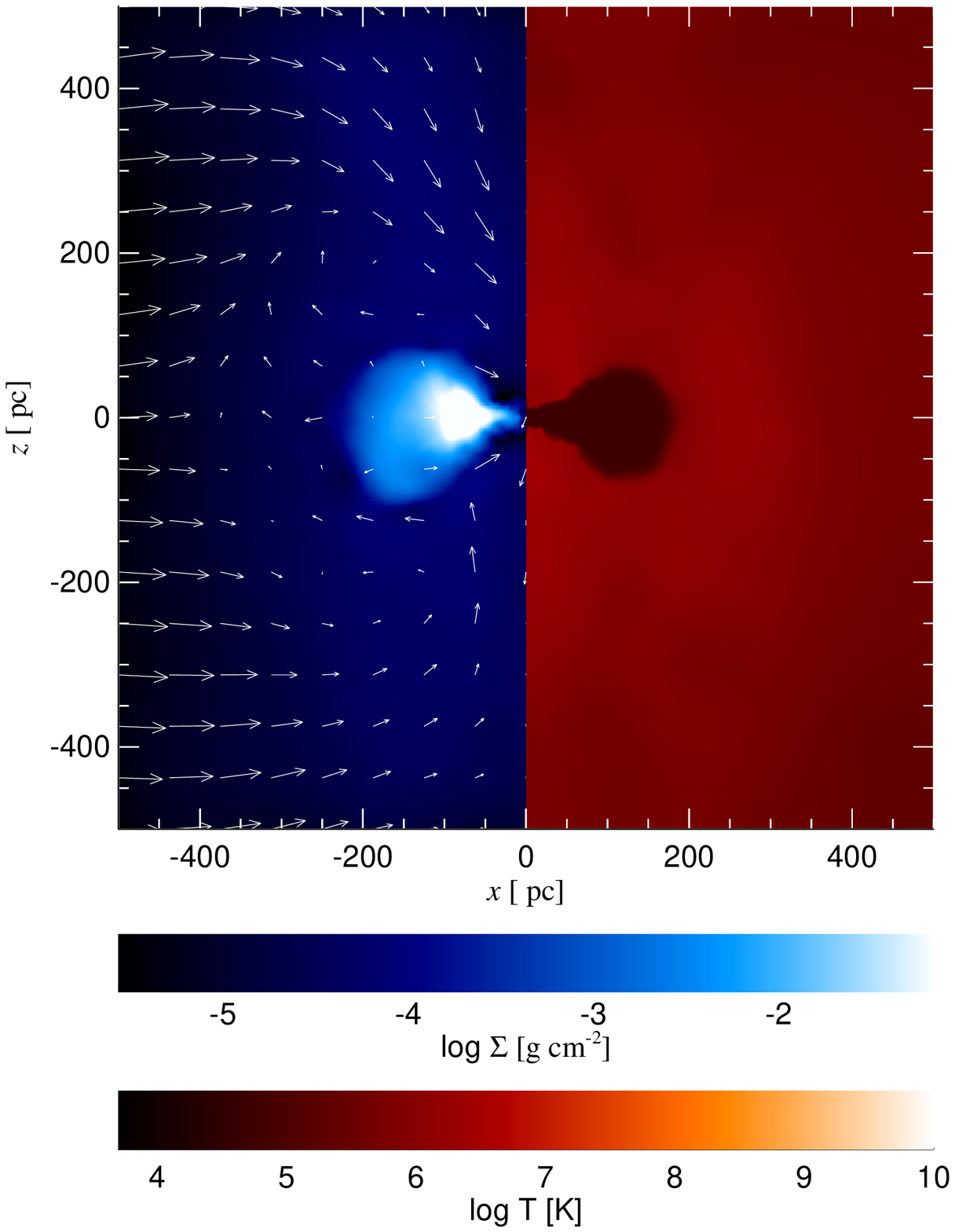}}
  \caption[Evolution of the central regions - side view]{{\bf Left:} Side view
    of gas surface density (left half of the panel, blue-white) and
    temperature (right, red-orange) in the central $2$~kpc of the `Base'
    simulation at $t = 1$~Myr. The CMZ has been enveloped by the external ISM
    shock fronts, but it still maintains the strong collimation of the diffuse
    cavities. The cavities are also filled with hot $10^8-10^9$~K gas, which
    allows them to expand in all directions once the feedback from \sgra\ has
    switched off. {\bf Right:} Central $500$~pc of the `Base' simulation at $t
    = 6$~Myr. The CMZ remains, although its density structure is
    perturbed. There is also a ``back flow'' of warm ($T \simeq T_{\rm vir}$)
    gas into the central regions, evacuated by the buoyant rise of the
    bubbles.}
  \label{fig:Base_early2}
\end{figure*}

Finally, there is a diffuse spherically symmetric gaseous ``halo'' extending
between $r_{\rm in} = 1$~pc and $r_{\rm out} = 15$~kpc with density following
the isothermal profile given by equation \ref{eq:rho}. The gas fraction
$f_{\rm g}$ is a free parameter of our model and is varied between $4 \times
10^{-4}$ and $4 \times10^{-3}$ in different tests, with the fiducial value of
$10^{-3}$ for the `Base' model (this corresponds to a particle density $6
\times 10^{-4}$~cm$^{-3}$ at $R = 5$~kpc; see the Discussion section for the
validity of this choice for gas density). The number of particles in the halo
is set by their mass (we use same SPH particle masses for the CMZ disc and the
halo). The gas temperature in the bulge halo is initially set to $T_{\rm halo}
= T_{\rm vir} = 2.5 \times 10^5$~K, which corresponds to the virial
temperature of the bulge. The initial gas velocity in the bulge is set to
zero.

This halo setup we use is probably oversimplified. For example, we do not
account for the likely anisotropy of the initial gas distribution of the
``halo'' due to overall rotation of the Galaxy. We address this point
qualitatively in the Discussion section.

\section{Results} \label{sec:results}

Table \ref{table:paramsnew} shows the list, parameters and main results of all
of the simulations that we present in this paper. Results of the three
simulations in the top of the Table, separated from the rest by a horizontal
line, appear to be a reasonable match to the {\it Fermi} Bubble observations
by \cite{Su2010ApJ}. The rest of the simulations produce bubbles that are
unlike the observed ones.

We first describe the evolution and properties of the `Base' model ($f_{\rm g}
= 10^{-3}, t_{\rm q} = 1$~Myr).

\subsection{Base simulation} \label{sec:Base}

\subsubsection{Small and intermediate scales}\label{sec:small}

Overall, the dynamics of gas in the simulation closely follows our analytical
expectations. As soon as the quasar switches on, the spherically symmetric
outflow hits both the CMZ and the halo gas. The left panel of Figure
\ref{fig:Base_early1} shows the face-on view of the CMZ disc at time
$t=0.1$~Myr. Note that only the central $\sim 25$ pc of the disc were
evacuated by outflow from \sgra. The right panel of the same figure shows the
edge-on projections of both gas column density and temperature (which we
present in a single panel because of the azimuthal symmetry of the gas
flow). The arrows show gas velocity vectors projected on the plane of the figure.

As expected, the spherical ``halo'' (the diffuse gas component) is affected by
the outflow much more than the CMZ, with the contact discontinuity between the
wind and the shocked ambient medium at a distance of about $300$~pc. We also
note that the bubbles do contain some gas mainly closer to the interface with
the upper layers of the CMZ. That gas is heated to temperatures above
$10^8$~K. There is a transition region between the almost-spherical
  outflow perpendicular to the Galactic plane and the stalled outflow against
  the CMZ, completing the figure-8 morphology of the whole flow.

As time proceeds the inner hole in the CMZ disc grows in size, engulfing most
of the disc by the end of the simulation (see \S \ref{sec:Base_centre} for a
fuller discussion of this). The cavities opened by the outflow in the
directions perpendicular to the Galactic plane grow even more. The left panel
of Figure \ref{fig:Base_early2} shows the edge-on view of the inner $2$~kpc of
the Galaxy at time $t=1$~Myr. One notes the hour-glass shape of the cavities
opened by the outflow, and a strong gradient in the outflow velocity with
angle $\theta$ ($\tan \theta = |z/x|$) meausured from the $z$-axis. In
particular, the maximum velocity is reached at $\theta=0$, such that velocity
there is consistent with the analytical prediction of $v_{\rm e} \sim
2000$~km/s, and the minimum is at $\theta=90^\circ$. There is also a
``failed'' outflow around $\theta\approx 65^\circ$; gas flowing along these
directions eventually falls on the Galactic plane, shadowed by the CMZ from
further \sgra\ feedback. Some material is ablated from the surfaces of the
CMZ. It expands and contributes to the tenuous gas filling the cavities.

The right panel of Figure \ref{fig:Base_early2} presents the edge-on view of
the central $500$~pc at the end of the simulation, at $t=6$~Myr. Due to the
relative buoyancy of the hot gas in the cavities with respect to the cooler
``ambient'' gas, the former leaves the region by that time, being replaced by
the latter. This ``back flow'' of warm ($T \simeq T_{\rm vir}$) gas returning
to the central region after \sgra\ switched off is clearly seen in the pattern
of velocity vectors on the left side of the right panel of Figure
\ref{fig:Base_early2}. We also point out that while the CMZ has been
significantly affected by the outflow from \sgra, most of it remains in the
region in the form of a ring discussed further in \S \ref{sec:Base_centre}.

\subsubsection{Large scales}\label{sec:large}

\begin{figure*}
  \centering
    \subfloat{\includegraphics[width=0.33\textwidth]{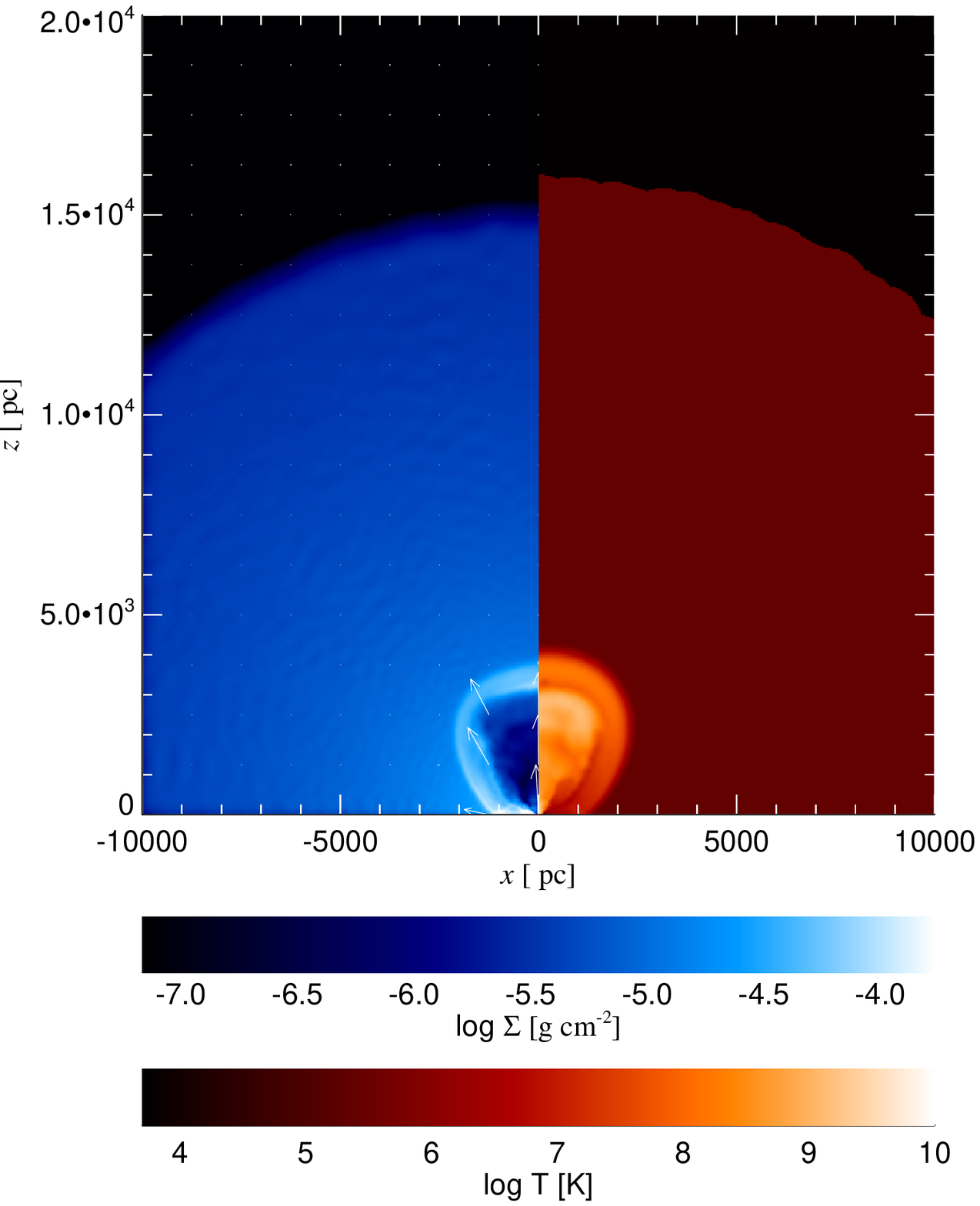}}
    \subfloat{\includegraphics[width=0.33\textwidth]{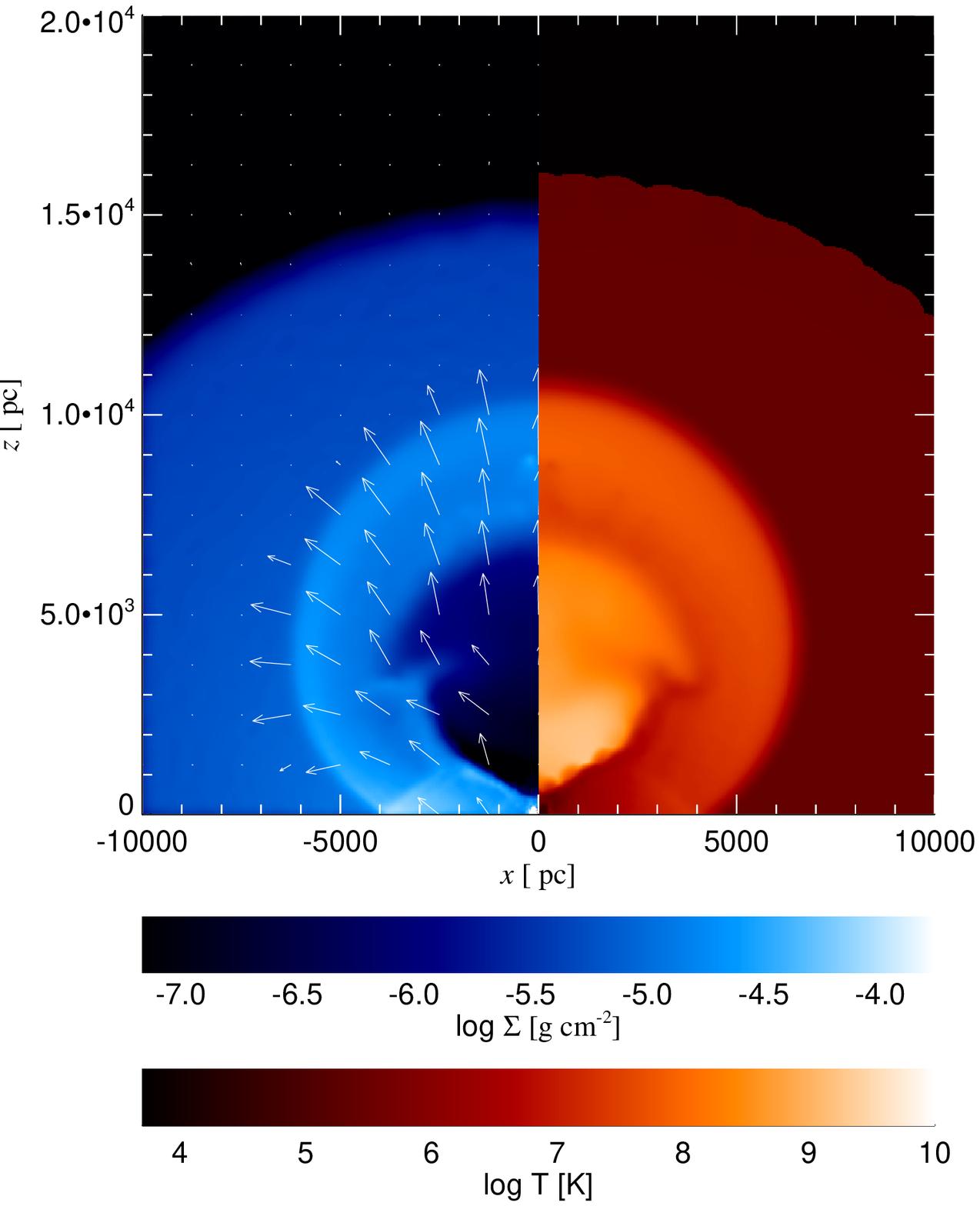}}
    \subfloat{\includegraphics[width=0.33\textwidth]{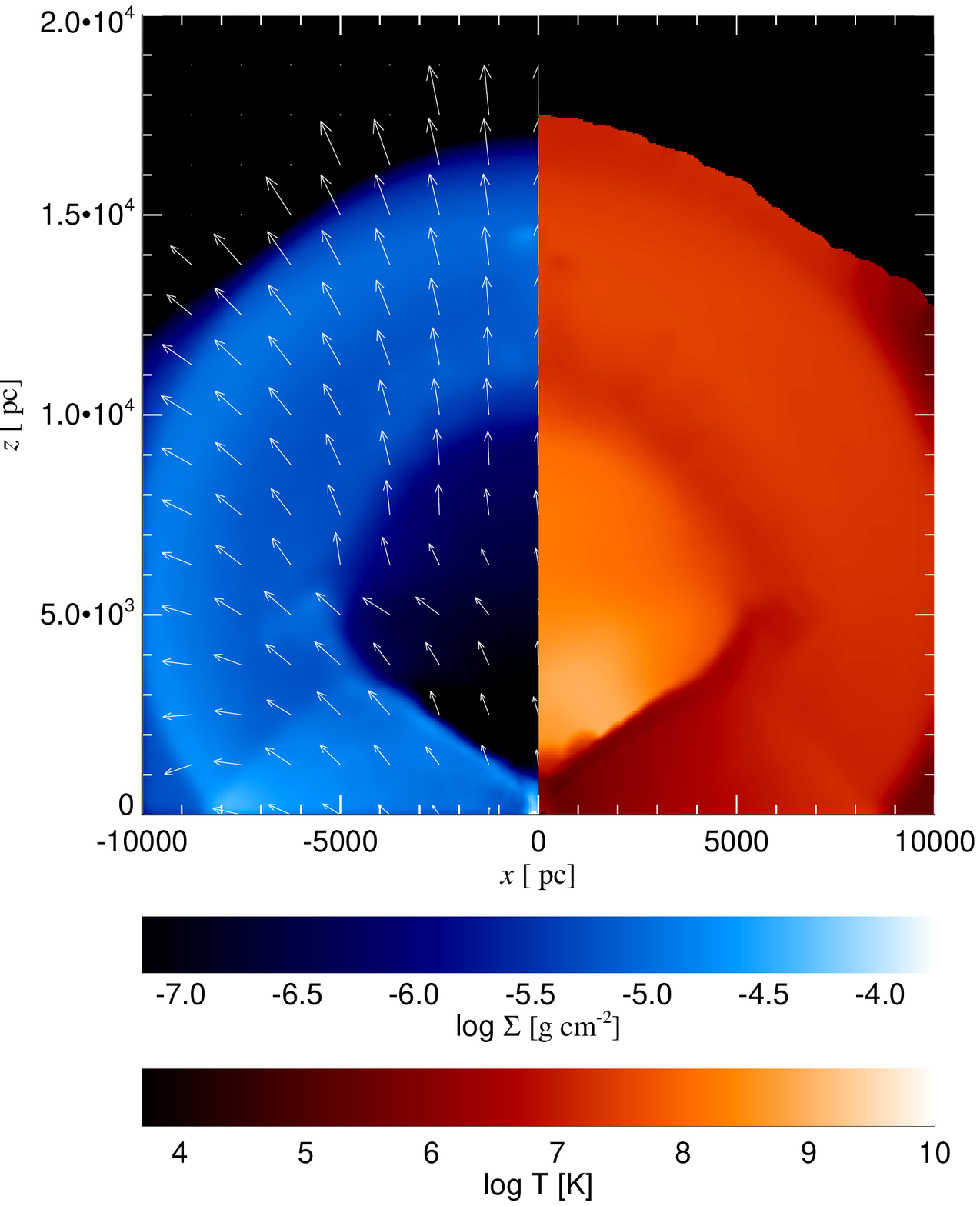}}
  \caption[Bubble evolution in `Base' simulation]{Gas evolution on large
    scales in the `Base' simulation; the three panels correspond to $t = 1, 3$
    and $6$~Myr, from left to right. Only the positive-z side of the
    computational domain is shown, due to symmetry around the Galactic
    plane. The CMZ strongly collimates the outflow and allows the formation of
    a teardrop-shaped cavity with a morphology very similar to that of the
    observed {\it Fermi} bubbles. The bubbles continue to expand and rise due
    to high pressure and low density, even once the feedback has switched
    off.}
  \label{fig:Base_evol}
\end{figure*}

\begin{figure*}
  \centering
    \subfloat{\includegraphics[width=0.33\textwidth]{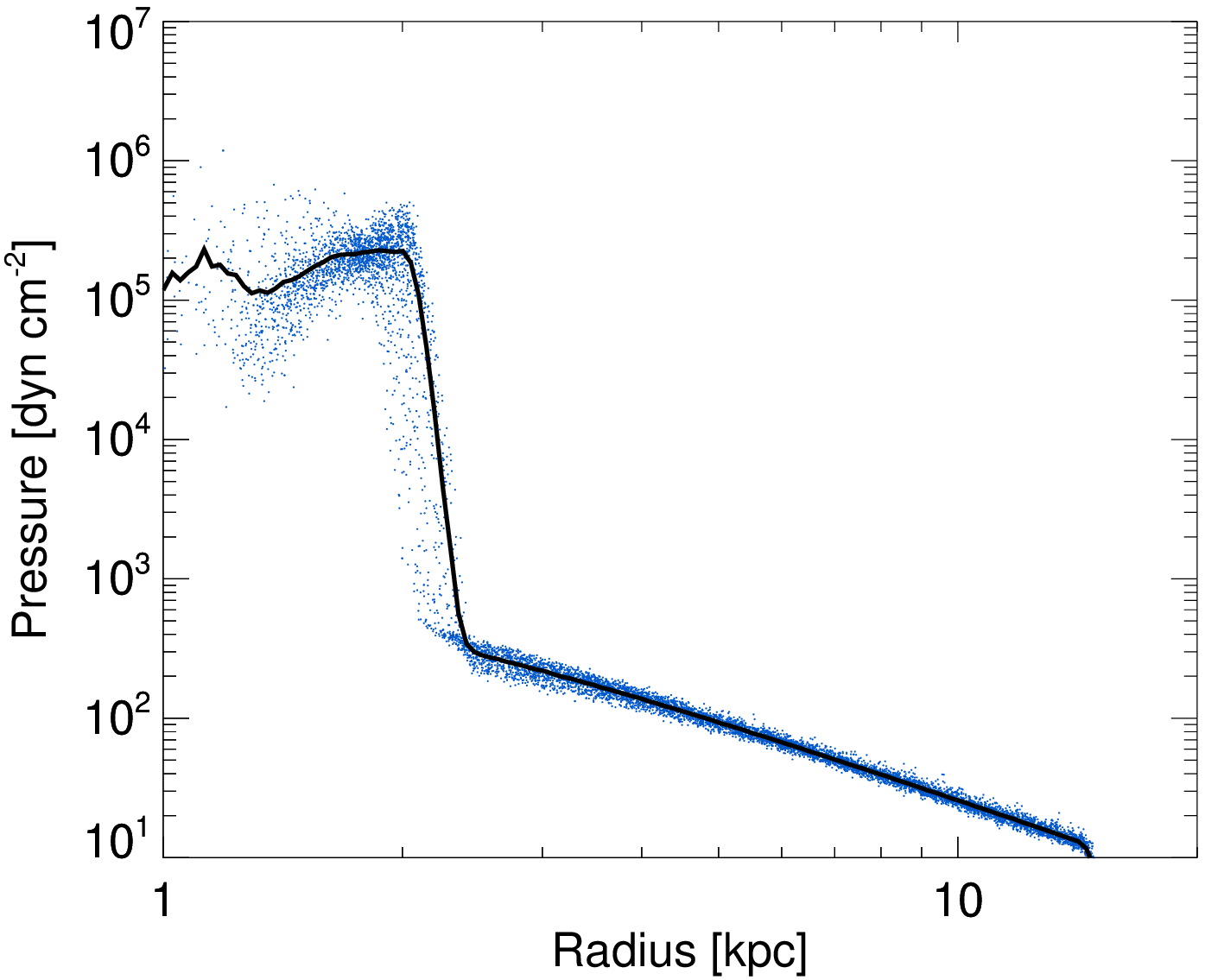}}
    \subfloat{\includegraphics[width=0.33\textwidth]{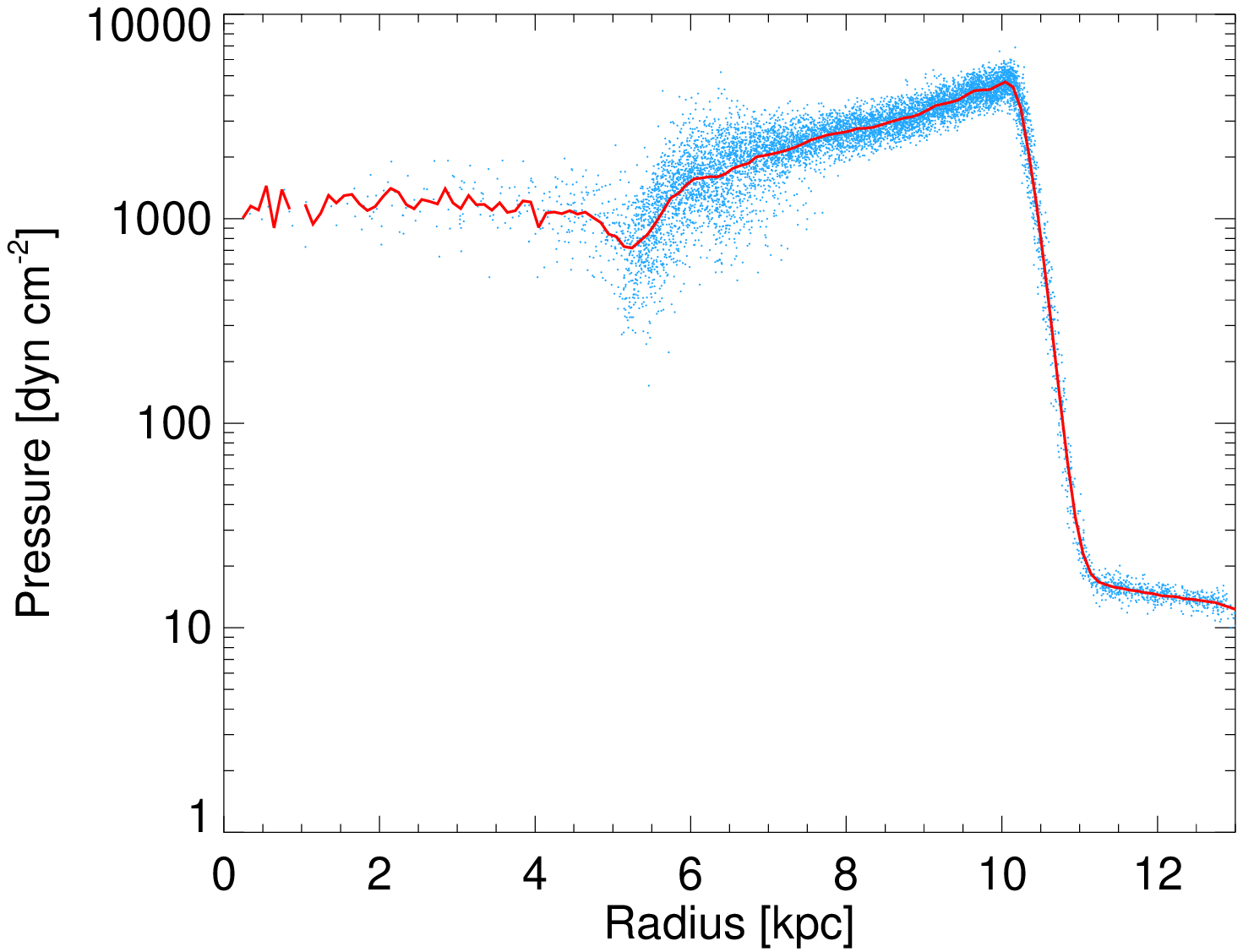}}
    \subfloat{\includegraphics[width=0.33\textwidth]{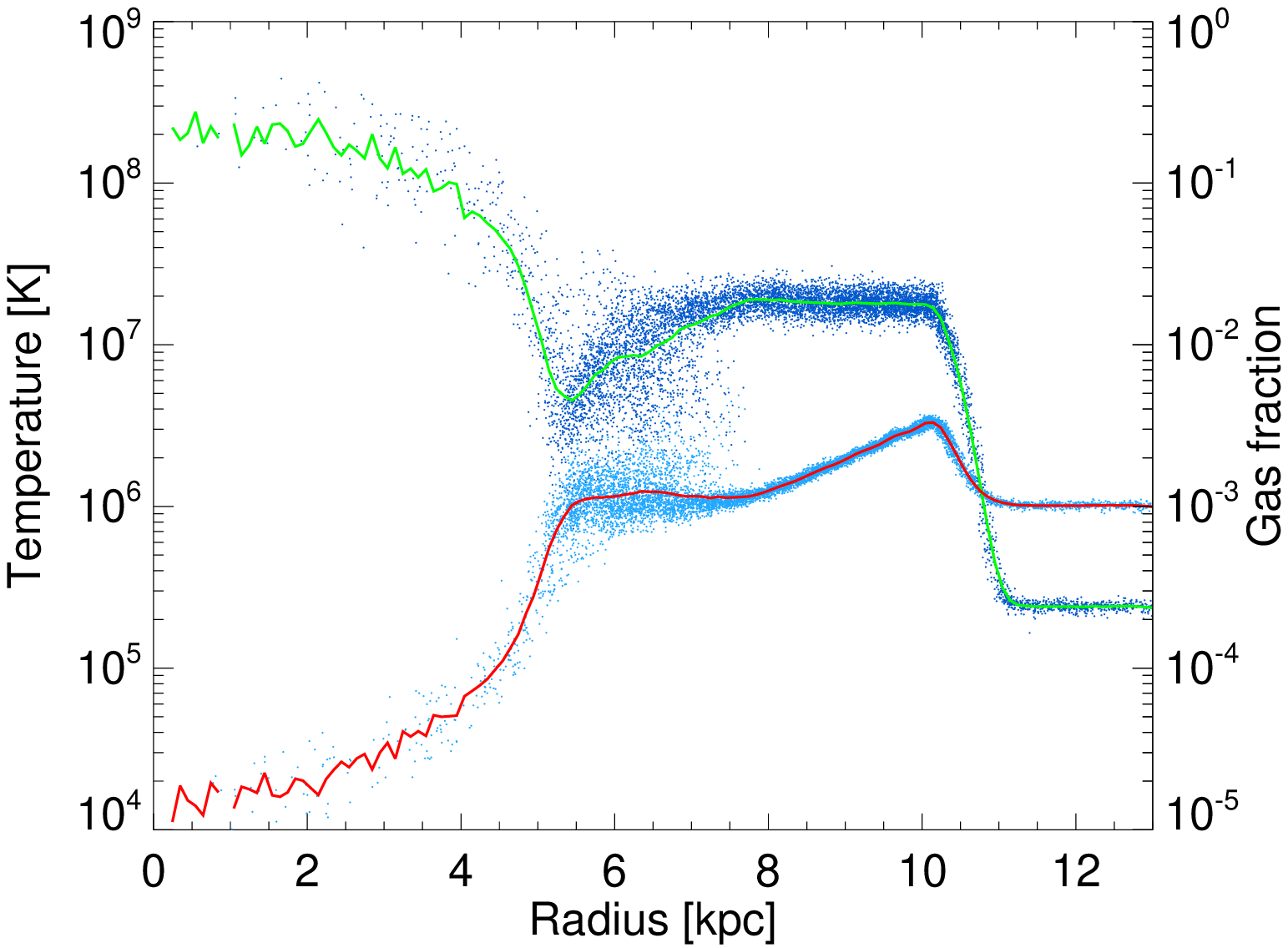}}
  \caption{{\bf Left}: Gas pressure as a function of cylindrical radius
    ($r_{\rm cyl}^2 = x^2 + y^2$) for a selection of SPH particles at $1 < |z|
    < 2.5$~kpc (middle of the bubble) for the `Base' simulation at $t =
    1$~Myr; the solid line is the mean value at each radius. The central
    cavity is significantly overpressurised in comparison to the ambient
    medium, leading to lateral expansion of the bubbles. {\bf Middle}: Same
    but at time $t = 6$~Myr when the bubble is much larger. The vertical cut
    chosen is now $4.5 < |z| < 6$~kpc. Note that the gas pressure has dropped
    significantly and is slowly varying accross the bubble and the shocked
    region.  {\bf Right}: Temperature (green curve and dark blue points, left
    scale) and gas fraction (red curve and light blue points, right scale)
    against cylindrical radius for the same time and particle cut as the
    middle panel. The hot diffuse inner cavity and dense surrounding medium
    are obvious and have rather sharp edges (thickness $\sim1$~kpc). Solid
    lines show mean values of gas fraction and temperature at each radius.}
  \label{fig:Base_prop}
\end{figure*}

We now discuss the larger scales of the ``Base'' simulation.  Figure
\ref{fig:Base_evol} shows the edge-on views of the simulation domain at times
$t = 1, 3$ and $6$~Myr (left, middle and right panels, respectively).  The
left panel in particular shows that by the time \sgra\ switches off the
cavities are still rather small on Galactic scales: their height is $R
\sim3$~kpc and the maximum width is $d \sim 2.5$~kpc.

Further evolution of the hot bubbles is driven by the inertia of the outflow,
the buoyancy of the bubbles and the fact that they are significantly
over-pressurized with respect to the ambient medium (cf. the left panel of
Figure \ref{fig:Base_prop}).  Bubble expansion proceeds in an almost
self-similar fashion, except for a ripple at roughly the middle of the bubble
height (see Fig. \ref{fig:Base_evol}, middle panel). It is most likely a
Kelvin-Helmholtz unstable mode which arises due to the material inside the
bubble moving parallel to the surface of the contact discontinuity. The ripple
rolls over and disappears by $t=6$~Myr (Fig. \ref{fig:Base_evol}, right).

\begin{figure}
  \centering
\includegraphics[width=0.48\textwidth]{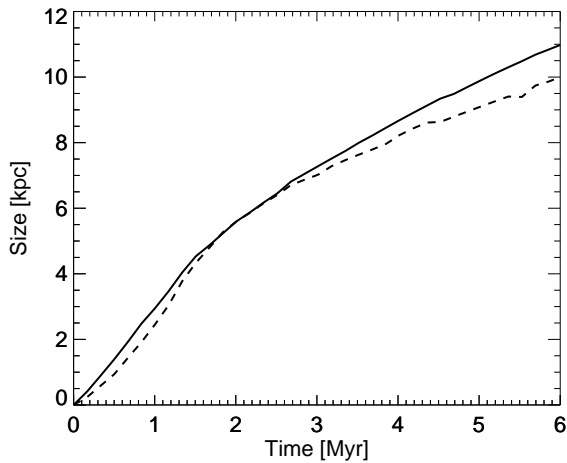}
  \caption{Height (solid line)
    and width (dashed line) of the bubbles as function of time in the `Base'
    simulation.}
  \label{fig:Base_rvst}
\end{figure}

By the end of the simulation, the cavities have expanded to reach a height of
$\sim 11.5$~kpc (Fig. \ref{fig:Base_evol}, right panel), consistent with the
observed vertical extent of the {\it Fermi} bubbles \citep{Su2010ApJ}. The
width, at $\sim 9$~kpc, is slightly larger than observed ($d_{\rm obs} \sim
6$~kpc); we return to this point in the Discussion section. The cavities are
filled with very hot ($T_{\rm bub} \sim 2\times10^8$~K $\sim17$~keV) and
diffuse ($f_{\rm g, bub} \sim 2\times10^{-5}$; see Fig. \ref{fig:Base_prop},
right) gas. For completeness, Figure \ref{fig:Base_rvst} shows the time
evolution of the bubble's height, $R$, and width, $d$. The lateral expansion
is initially somewhat faster than the vertical, possibly because some of the
material expanding outward along the disc plane was blown off the top of the
CMZ disc (rotating gas is easier to blow away due to centrifugal
force). As is seen in the figure, in the next few Myr, the bubbles are as wide
as they are tall at $t \sim 2$~Myr.  Subsequently, vertical expansion becomes
faster than lateral one. This may be driven by buoyancy: as the bubbles
continue to rise up, their lower edges ``lift off'' from the plane, and the
cooler gas can start flowing back along the plane.

The bubbles are slightly detached from the very centre of the Galaxy; the
gradual increase in this detachment is visible in Figure \ref{fig:Base_evol},
and also in the right panel of figure \ref{fig:Base_early2}.  This detachment
is caused by the bubbles rising buoyantly out of the Galaxy potential. By $t =
6$~Myr, the gap between the centre and the lower edges of the bubbles is $h
\sim1$~kpc, in agreement with observations that permit any value of the gap
size below $\sim2$~kpc \citep{Su2010ApJ}.

The left panel of Figure \ref{fig:Base_prop} shows the SPH particle pressure
versus their cylindrical radius defined as $r_{\rm cyl}^2 = x^2 + y^2$
selected at a slab of gas with $1 < |z| < 2.5$~kpc at $t = 1$~Myr, roughly
corresponding to the midplane (i.e. half-height) of the bubbles at that
age. We observe that the difference in pressure within the bubble and outside
is greater than two orders of magnitude.  This excess pressure continues to
drive the expansion in directions both parallel and perpendicular to the
Galactic plane long after the quasar has switched off.

The middle panel of figure \ref{fig:Base_prop} presents the same as the left
figure but at the end of the simulation and for a slab of material at a larger
$|z|$ ($4.5 < |z| < 6$~kpc, corresponding roughly to the middle of the bubble
that has now risen further from the Galactic plane at $t = 6$~Myr). Note that
by $t = 6$~Myr, the gas pressure inside the bubble has dropped by two orders
of magnitude; the pressure of the surrounding ISM is also lower, but only by
about a factor $10$, so the pressure difference between the bubble and its
surroundings is much smaller. Consequently, the bubble expansion is much
slower, as seen in Fig. \ref{fig:Base_rvst} and the right panel of
Fig. \ref{fig:Base_evol}. The vertical expansion of the bubbles persists for
longer due to residual momentum in the gas and buoyancy (see the last two
columns in Table \ref{table:paramsnew}).

\del{drops significantly and is about equal to that of
the surrounding medium. Therefore, lateral expansion is much slower than
vertical one by that time (see the velocity vectors in
Fig. \ref{fig:Base_evol}, right). The latter may still persist for some time
due to residual momentum in the bubble gas (see the last two columns in Table
\ref{table:paramsnew}).}

\subsubsection{Feedback effects on the CMZ}\label{sec:Base_centre}

\begin{figure*}
  \centering
    \subfloat{\includegraphics[width=0.33\textwidth]{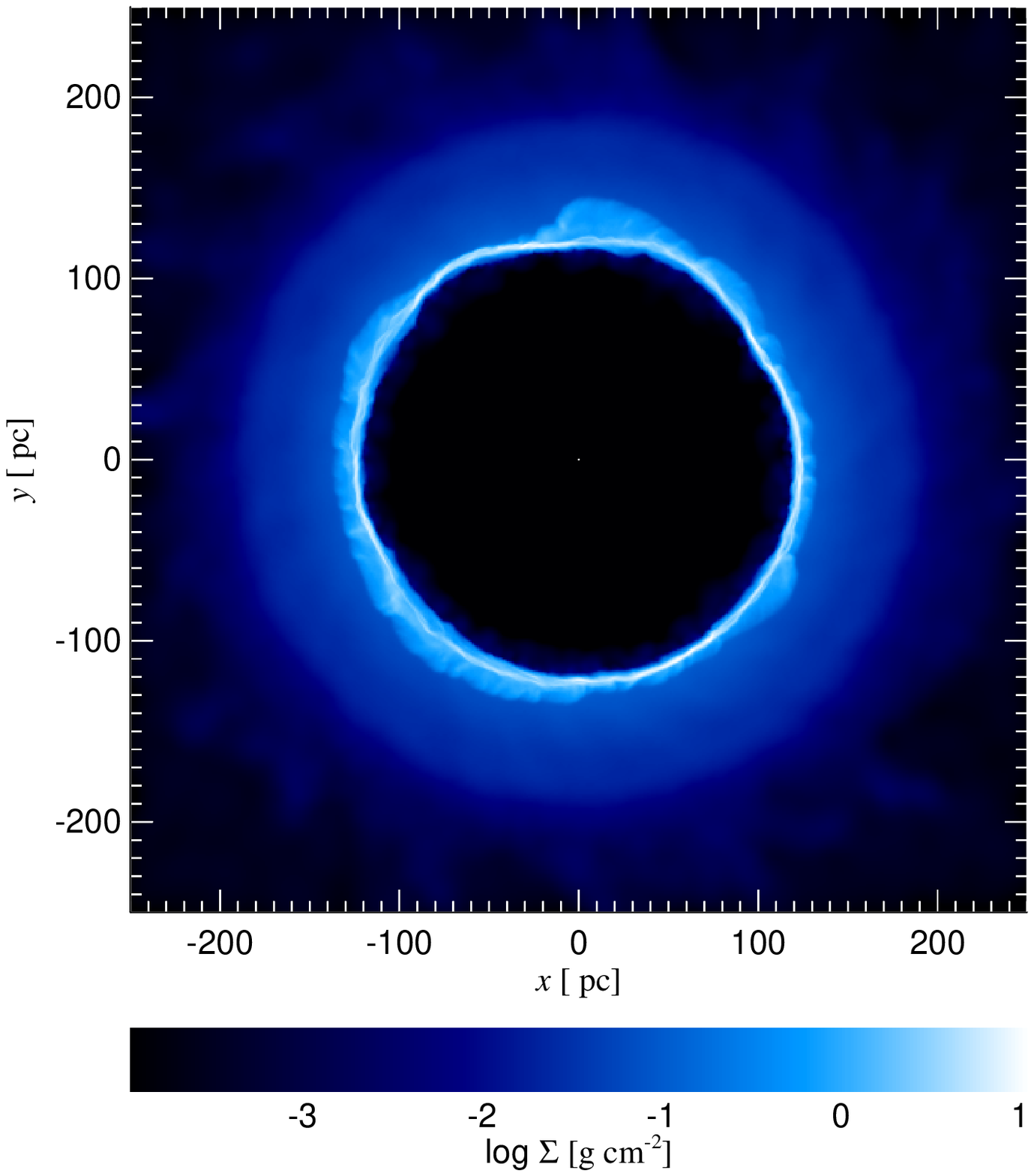}}
    \subfloat{\includegraphics[width=0.33\textwidth]{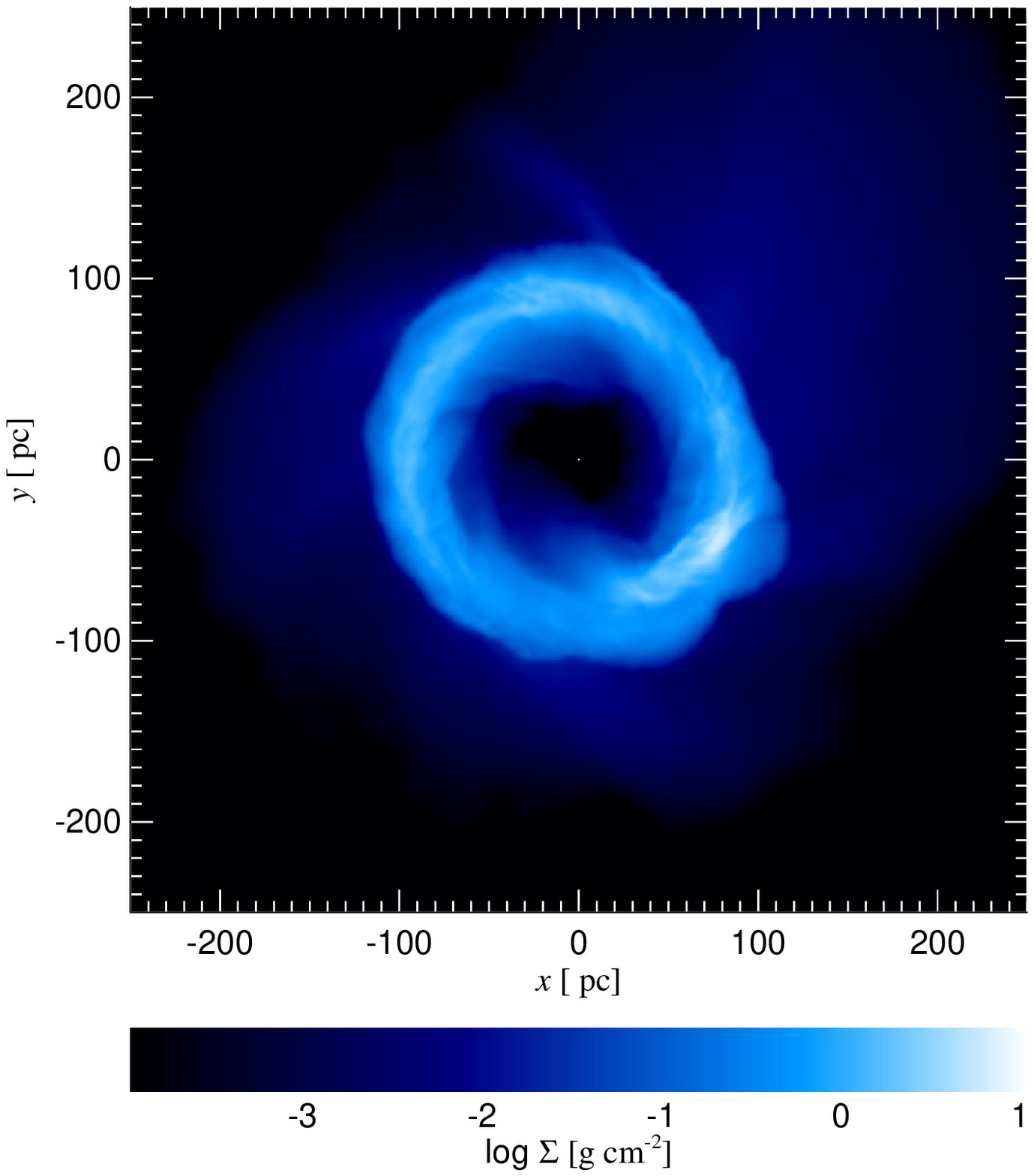}}
    \subfloat{\includegraphics[width=0.33\textwidth]{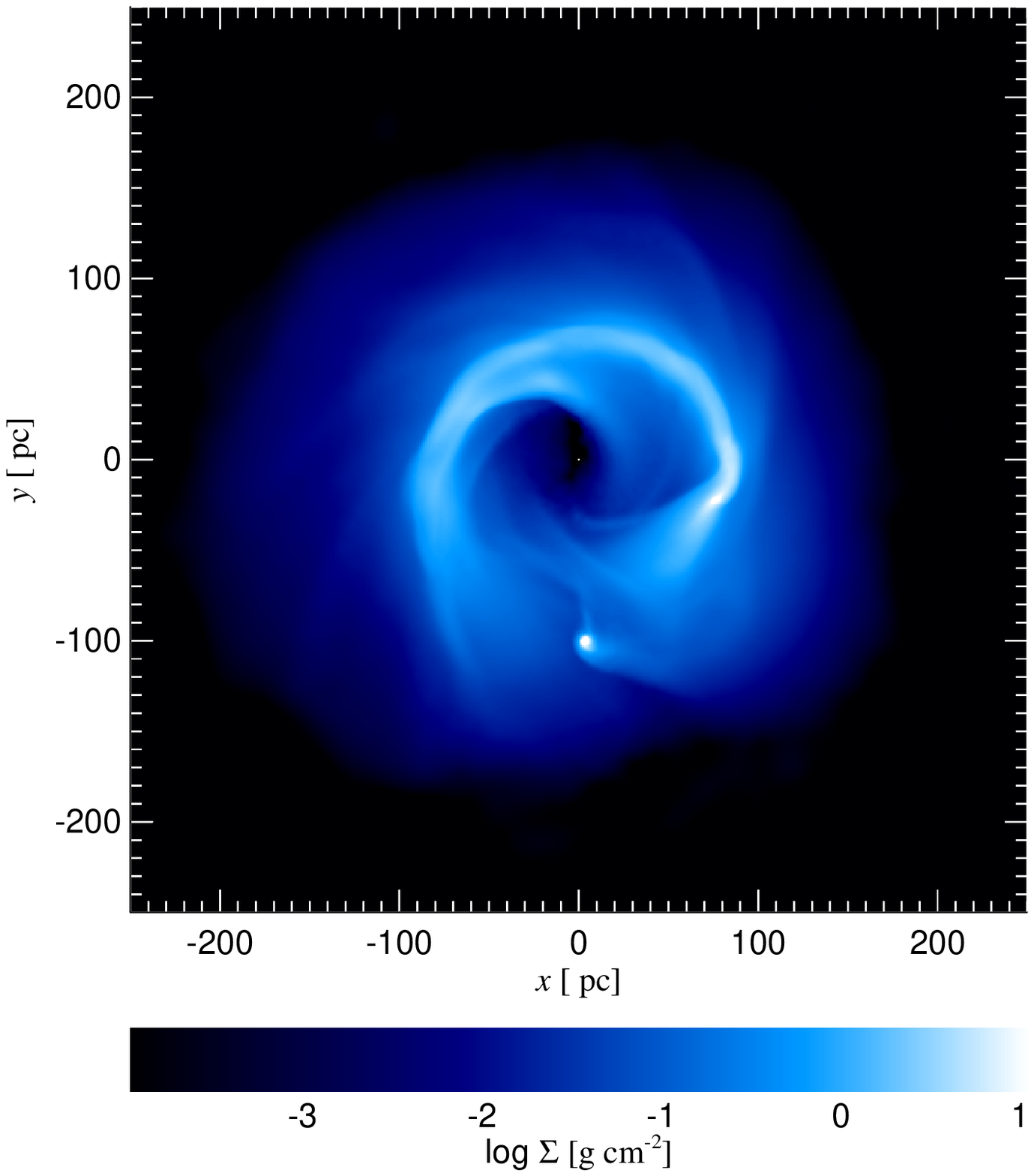}}
  \caption[Evolution of the central regions - face-on view]{Density of gas in
    the central $250$~pc of the `Base' simulation at $t = 1, 3$ and $6$~Myr
    (left, middle and right panels, respectively), as seen from above the
    Galactic plane. The inner regions of the CMZ are strongly affected by the
    quasar activity, forming a ring (left panel) which persists for several
    Myr (middle panel) while slowly spreading radially and developing unstable
    spiral filaments (right panel). A self-gravitating clump forms at $t =
    3.5$~Myr, with another about to form by the end of the simulation. The
    total mass of the CMZ remains almost constant throughout the simulation.}
  \label{fig:Base_centre}
\end{figure*}

In the simulation, we find that the CMZ is not dispersed by the outflow, a
result consistent with the prediction of the analytical argument (cf. Section
\ref{sec:waist}). However, the quasar wind is powerful enough to displace the
inner $\sim120$~pc of the CMZ to larger radii, resulting in formation of a
dense thin ring (Figure \ref{fig:Base_centre}, left).

The average radial expansion velocity of the inner parts of the CMZ,
$v_{\rm r,cmz} \sim 100$~km/s, is larger than the sound speed in  its gas
($c_{\rm s} \sim 20-40$~km/s, for $H/R = 0.125$ and $0.25$ respectively) and
comparable to its rotational velocity. Therefore some of the CMZ gas is
shock-heated to $T \sim 10^6$~K and expands vertically. This allows the quasar
wind to ablate the outer surfaces of the CMZ further. Gas from these regions
fills the voids in the halo (cf. Sections \ref{sec:small} and \ref{sec:large},
above). This process, however, removes only a small amount of mass: by the end
of the simulation, the CMZ mass (defined as the gas mass within a radius
$250$~pc in the Galactic plane and within $z \pm 100$~pc) has decreased by
less than $5\%$.

We would like to note another more significant dynamical effect of the
\sgra\ outflow on the CMZ that re-shapes the initial disc configuration into a
ring-like one. The outward force and pressure of quasar wind cause a
significant radial mixing in the CMZ. For example, some of the CMZ gas on the
inner face of the disc is sent outward by the wind just above the surface of
the outer regions of the disc, but stalls later on and falls back on the CMZ
in the outer regions. This failed outflow has a small angular momentum in
comparison to gas in the outer disc. The angular momentum of the CMZ gas is
then well mixed up. This establishes a narrow distribution of specific angular
momentum -- a ring. We note that this mechanism of ring formation is similar
in spirit to that found by \cite{Hobbs2011MNRAS} in their simulation S30,
although there shocks between material with different angular momentum were
due to initial conditions in the collapsing gas shell rather than quasar
feedback.

Subsequently, the ring slowly relaxes and spreads back somewhat into a disc
configuration due to viscous stresses; however, the viscous timescale for this
to happen at $R = 100$~pc is $t_{\rm visc} \sim 10^8
\left(0.1/\alpha\right)$~yr for $H/R = 0.25$, where $\alpha$ is the standard
\citet{Shakura1973A&A} viscosity parameter. This is much longer than the
timescales we are interested in, and thus unsurprisingly the ring persists to
the end of the simulation (Figure \ref{fig:Base_centre}, right).

Since the disc is only marginally gravitationally stable at the start of the
simulation, it comes as no real surprise that the ring becomes more unstable
than the original disc. As a result, some spiral density waves are visible in
the middle panel of Fig. \ref{fig:Base_centre}. Furthermore, a very massive
dense clump forms at $t\sim3.5$~Myr and there are hints of another forming by
the end of the simulation at a position $\lbrace x,y \rbrace \sim \lbrace
80,-20 \rbrace$~pc. The radius of the clump is $r_{\rm cl} \sim 5$~pc and its
mass $m_{\rm cl} \sim 10^7 \msun$. Its density is then $n_{\rm cl} \sim
7\times10^5$~cm$^{-3}$, $\sim 500$ times greater than the background potential
density at the clump's radial distance $R \simeq 100$~pc and similar to that
of dense star forming molecular cloud cores. Due to the adopted temperature
floor and lack of resolution in the simulations, the cloud cannot fragment
into smaller globules, but we obviously expect such an object to be unstable
to gravitational collapse. The result would presumably be a massive star
cluster. The orbit of the clump around \sgra\ is mildly eccentric ($e \sim
0.2$). This may be interesting as a potential route to formation of the Arches
cluster which has a rather non-circular orbit \citep{StolteEtal08} in
particular, but it may be also relevant to the origin of other young Galactic
Centre star clusters and GMCs (cf. further discussion in \S
\ref{sec:induced}).

\subsection{Dependence on the ambient gas density} \label{sec:fgdep}

\begin{figure*}
  \centering
    \subfloat{\includegraphics[width=0.33\textwidth, trim = 0 0 0 0, clip]{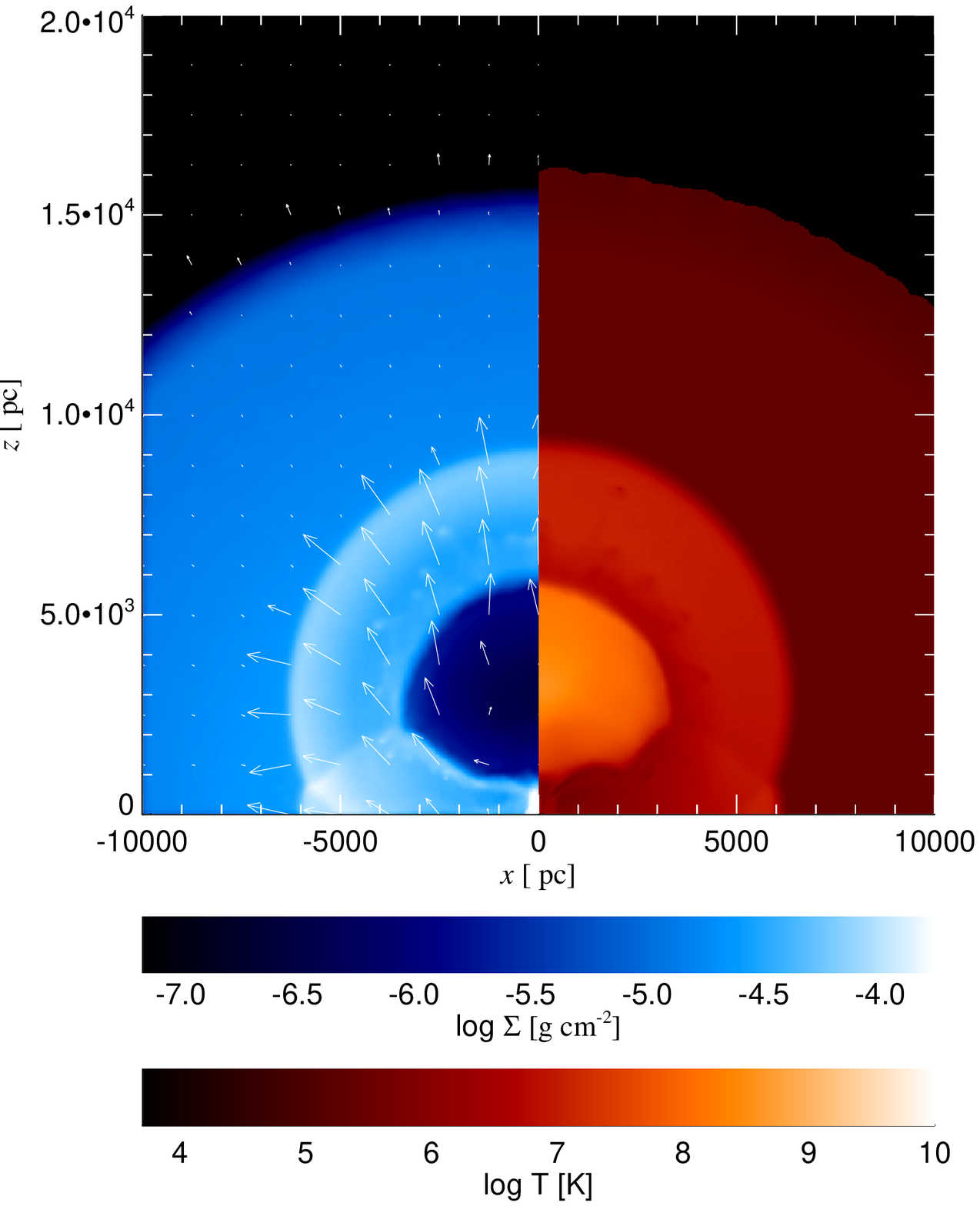}}
    \subfloat{\includegraphics[width=0.33\textwidth, trim = 0 0 0 0, clip]{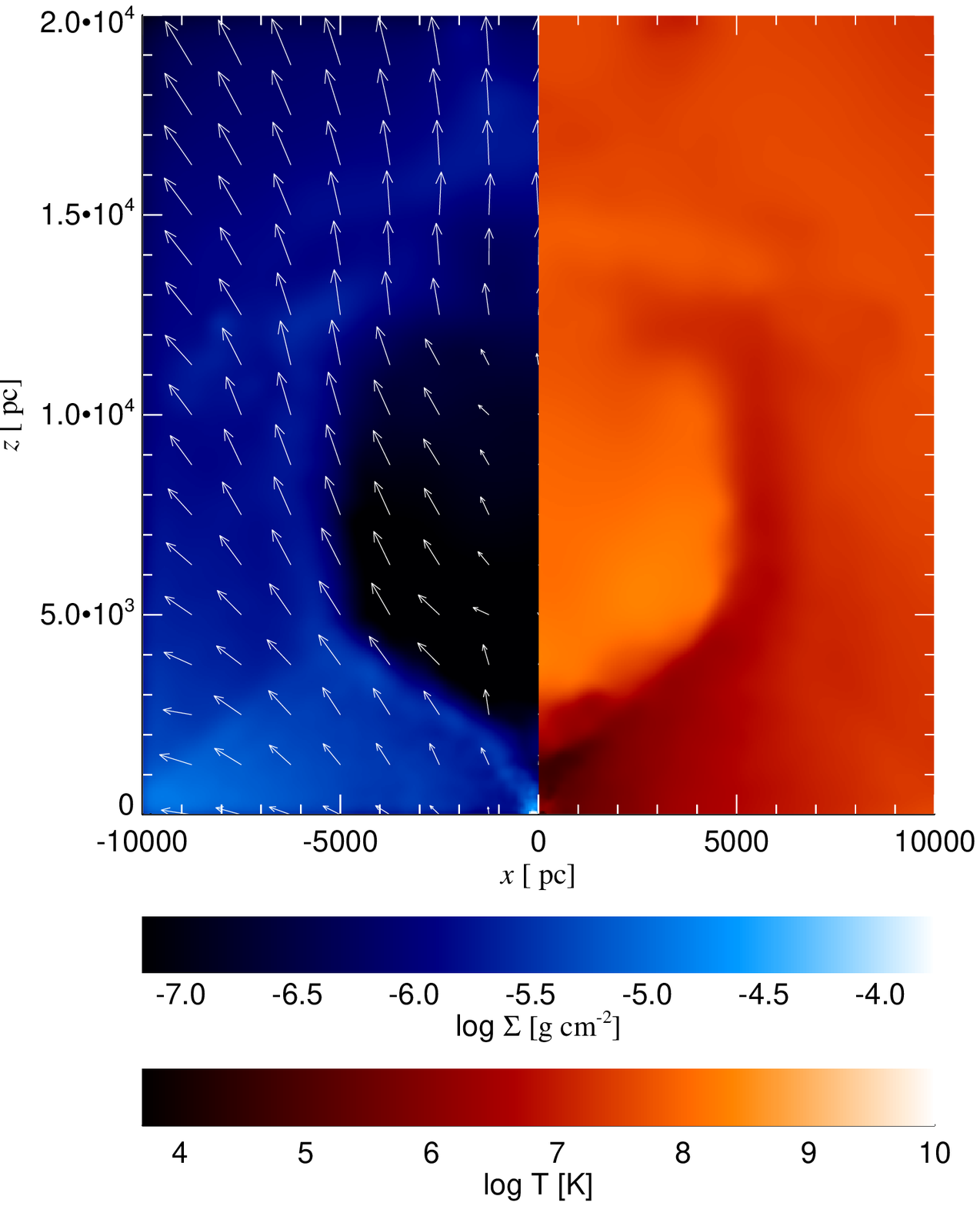}}
    \subfloat{\includegraphics[width=0.33\textwidth, trim = 0 0 0 0, clip]{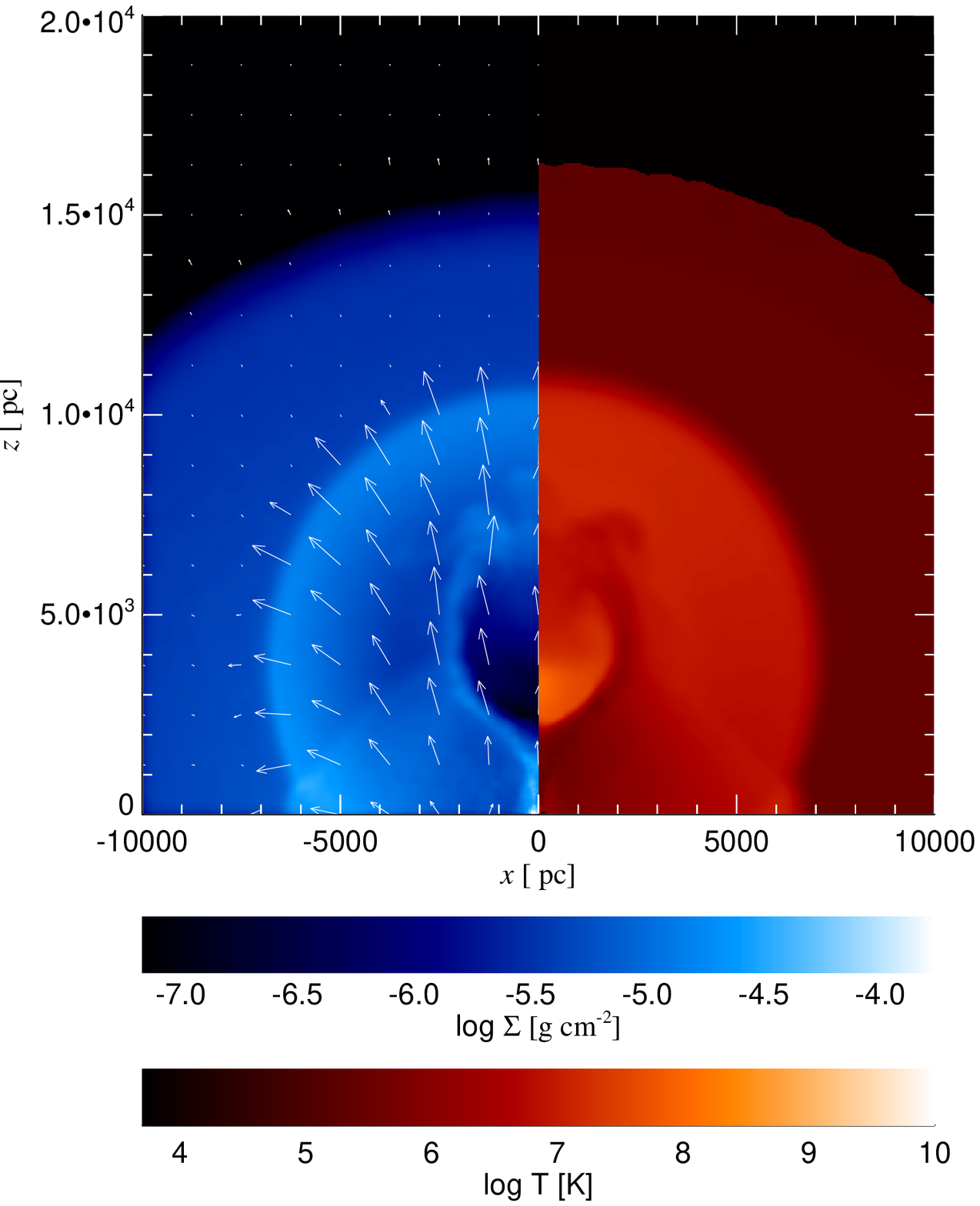}}
  \caption[Effects of variations in simulation parameters]{Density and
    temperature plots at $t = 6$~Myr of simulations with varying parameters.
    {\bf Left}: higher gas fraction $f_{\rm g} = 4 \times 10^{-3}$
    (`Fg-high'). {\bf Middle}: lower gas fraction $f_{\rm g} = 4 \times
    10^{-4}$ (`Fg-low'). {\bf Right}: lower quasar outburst duration $t_{\rm
      q} = 0.3$~Myr (`Tq-low').}
  \label{fig:vartests1}
\end{figure*}

The ambient gas density before \sgra\ quasar outburst is a free parameter of
the model. Therefore, we varied $f_g$ to see how our conclusions depend on
this paerameter. A fourfold increase in $f_{\rm g}$ (Simulation `Fg-high',
Fig. \ref{fig:vartests1}, left) results in a reduction of the bubble height by
$\sim40\%$ for the same $t_{\rm q}$. This is consistent with the analytical
prediction from equation (\ref{rstall}) and can be understood in terms of the
larger gas weight, which for an isothermal distribution is independent of
radius and is
\be 
W_{\rm g} = \frac{4 f_{\rm g} \sigma^4}{G} 
\ee
\citep{King2003ApJ, King2010MNRASa}. Denser gas weighs more, therefore
requiring more energy to be lifted to the same height. Since the energy input
is the same in both cases, higher values of $f_{\rm g}$ result in lower height
reached by the bubbles. In fact, in simulation `Fg-high', formally, the
stalling radius of the bubble is $R_{\rm stall} \sim10$~kpc, similar to the
size of the observed bubbles. However, the stalling time is $t_{\rm stall}
\sim50$~Myr, which is much greater than the time since the hypothesized
\sgra\ outburst. This shows that relatively high values of $f_{\rm g}$ are
definitely disfavored within our model of a recent outburst origin for the
{\it Fermi} bubbles (unless perhaps $t_{\rm q}$ is much longer than a 1 Myr; 
  but see Section \ref{sec:duration}).

Conversely, a simulation with a lower gas density initial condition (`Fg-low',
Fig. \ref{fig:vartests1}, middle) produces a bubble that is $\sim40\%$ taller,
but its interior temperature is rather low, blurring the distinction between
the bubble and its surroundings. Furthermore, the lower density bubble rises
further from the Galactic centre, producing a detachment $h_{\rm b} > 2$~kpc,
inconsistent with the data \citep{Su2010ApJ}. These disagreements allow us to
exclude the possibility of a significantly lower gas fraction in the halo as
well.

Our simulations show that the width of the bubbles, on the other hand, is
almost independent of the ambient gas density. This is probably because the
lateral expansion of the bubble is governed not only by the thrust from the
outflow, but also by the pressure balance on both sides of the bubble
(Fig. \ref{fig:Base_prop}, left and middle panels). The ambient gas pressure
increases with higher $f_{\rm g}$, but so does the pressure inside the cavity
due to lower bubble volume. The two changes compensate for each other, and the
pressure inside the bubbles remains several orders of magnitude greater than
in the external medium, leading to lateral expansion. Additionally, this
expansion happens on an approximately dynamical timescale, which is given by
the properties of the background potential and is also independent of $f_{\rm
  g}$.

\begin{figure*}
  \centering
    \subfloat{\includegraphics[width=0.33\textwidth, trim = 0 0 0 0, clip]{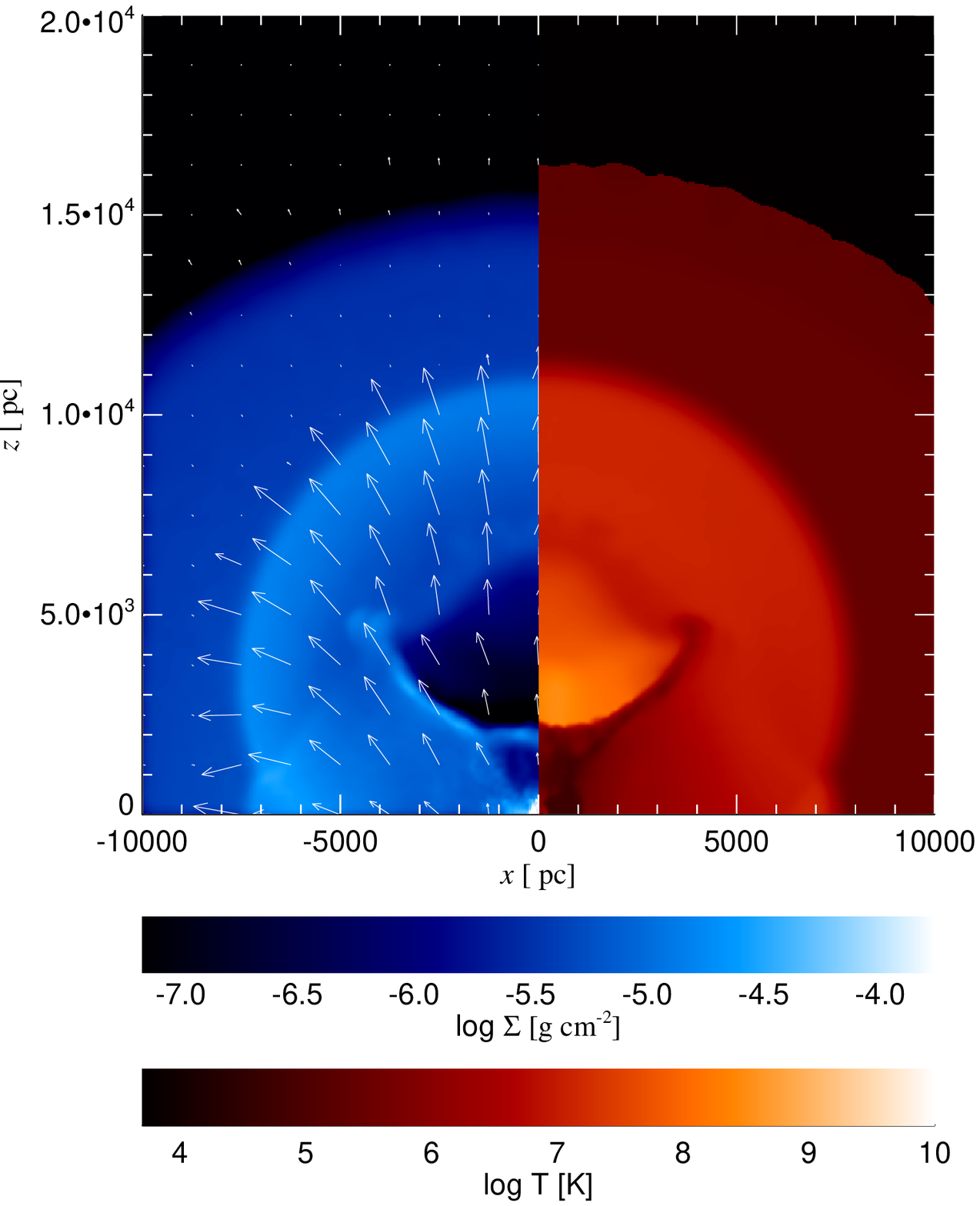}}
    \subfloat{\includegraphics[width=0.33\textwidth, trim = 0 0 0 0, clip]{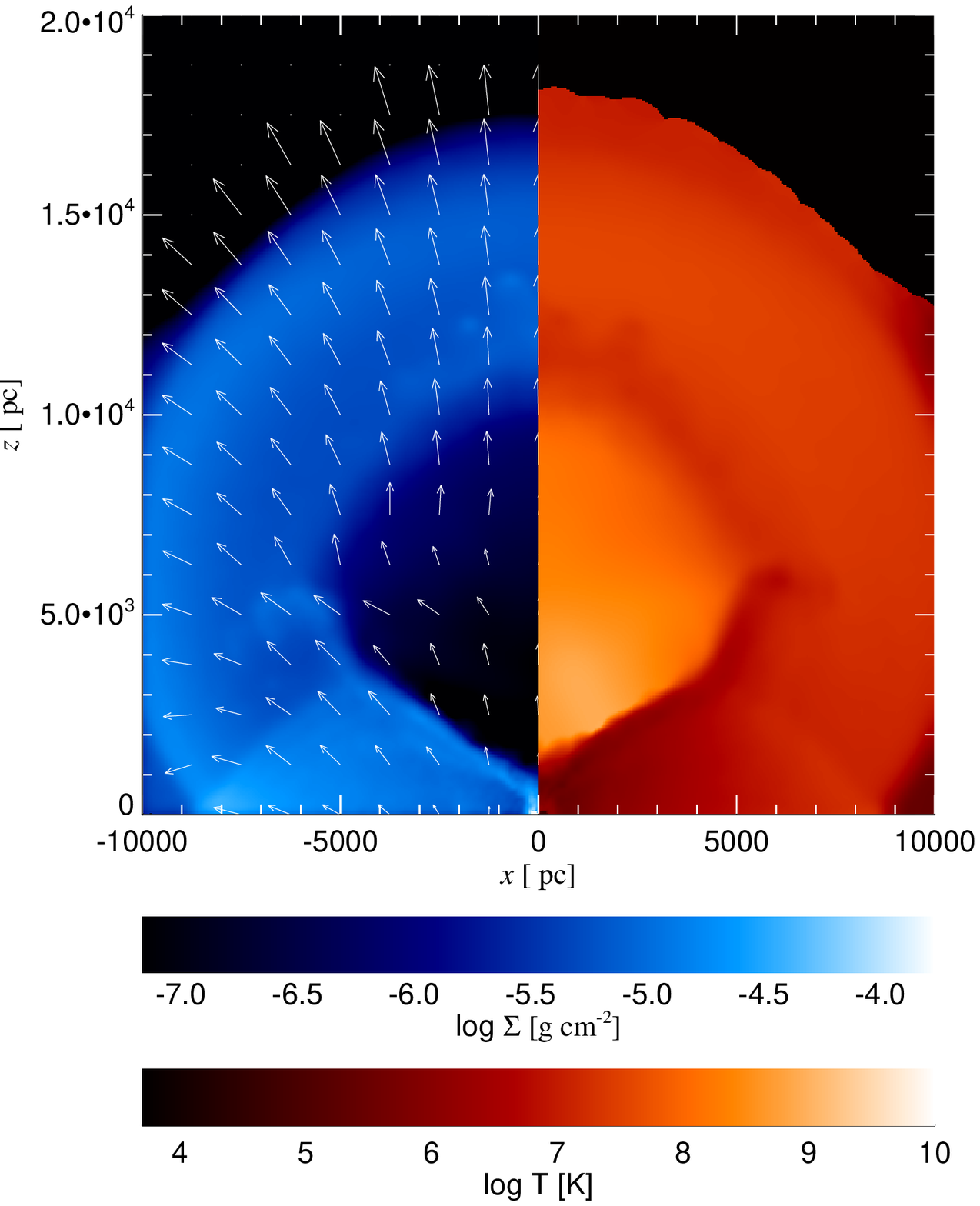}}
    \subfloat{\includegraphics[width=0.33\textwidth, trim = 0 0 0 0, clip]{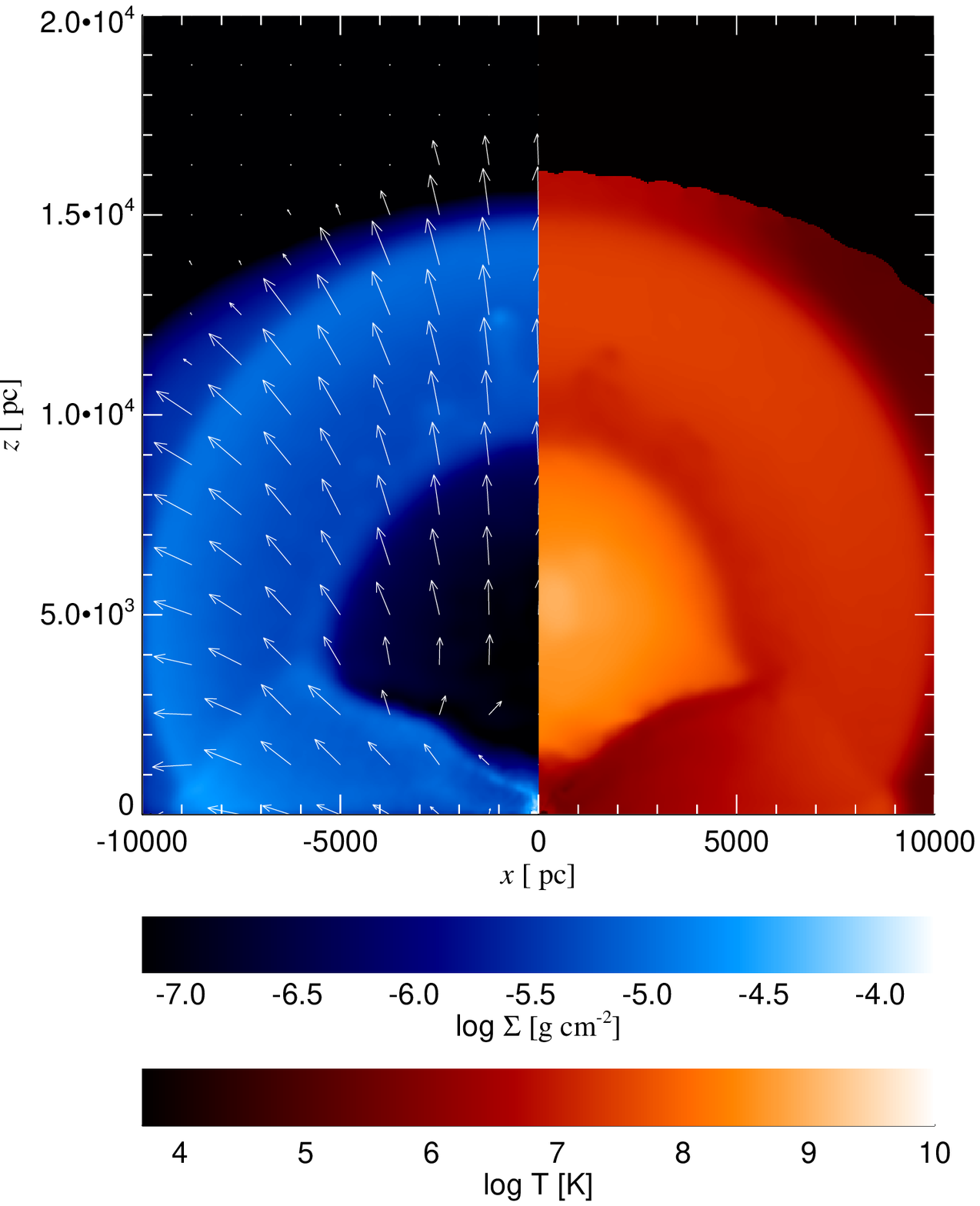}}
  \caption[Effects of variations in simulation parameters]{Density and
    temperature plots at $t = 6$~Myr of simulations with varying parameters.
    {\bf Left}: CMZ mass decreased by a factor 10 (`Mc-low'). {\bf Middle}:
    CMZ aspect ratio decreased to $H/R = 1/8$ (`Hr-low'). {\bf Right}: a
    different heating-cooling prescription has been used (''Cool').}
  \label{fig:vartests2}
\end{figure*}

\subsection{Dependence on the outburst duration} \label{sec:duration}

We now explore how simulation results depend on the outburst duration.
Simulation `Tq-low' (see Table 1) is identical to the `Base' simulation except
$t_{\rm q} = 0.3$~Myr.  As expected, lower $t_{\rm q}$ results in physically
smaller bubbles (see Fig. \ref{fig:vartests1}, right panel). Due to a smaller
amount of energy injected and a longer duration of adiabatic expansion phase
after quasar switch off, the contrast between the temperature and density in
the bubble interior and the surrounding shell is also smaller. The bubbles do
have lower densities (effective $f_{\rm g} \sim 4 \times 10^{-4}$, compared
with the ambient $f_{\rm g} \sim 10^{-3}$) and somewhat higher temperatures
($T_{\rm bub} \sim 10^7$~K) than the surroundings, and thus do rise buoyantly
away from the centre, resulting in large ($\gtrsim3$~kpc by $6$~Myr)
separations of the bubbles from the centre. Vertical expansion of the bubble,
however, is slower than in the `Base' simulation; the outer edges of the
bubbles reach only $\sim5-6$~kpc by the end of the simulation. In addition,
the bubble width is significantly smaller than in the `Base' simulation, since
a lower energy input results in a lower pressure inside the bubble, therefore
lateral expansion is also slower.

This result provides a rough lower limit to the quasar outburst duration required to
inflate the bubbles: $t_{\rm q} > 0.3$~Myr, and quite likely $t_{\rm q}
\sim1$~Myr. This is compatible with the analytical results, where we found
$t_{\rm q} > 2.5 \times 10^5$~yr from morphological arguments. We discuss this
point further in the Discussion section.

\begin{figure}
  \centering
\includegraphics[width=0.48\textwidth]{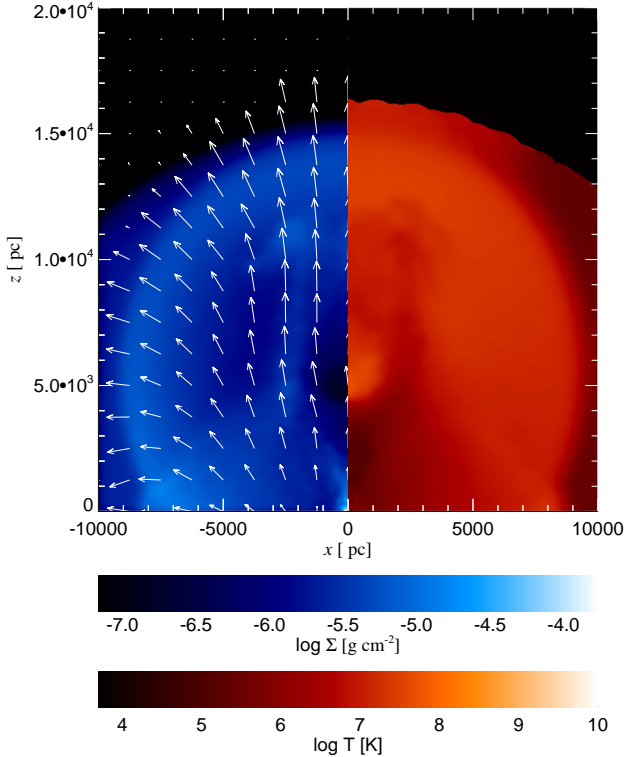}
  \caption{The structure of the bubbles at the end of the simulation
    `Both-low' in which both the duration and background density of gas are
    reduced from that of the `Base' simulation to $t_{\rm q} = 0.3$~Myr and
    $f_{\rm g} = 4 \times 10^{-4}$. The morphology of the bubble produced in
    this simulation is inconsistent with the observations.}
  \label{fig:both_low}
\end{figure}

Given that a lower value of $t_{\rm q}$ produces smaller bubbles, while a
lower value of $f_{\rm g}$ increases their size (cf. Section \ref{sec:fgdep}),
 is it possible that reducing both $t_{\rm q}$ and $f_{\rm g}$ may yield
as good or better morphological fit to the observed bubbles as the ``Base''
simulation? We run a simulation to test this idea (see simulation ``Both-low''
in Table \ref{table:paramsnew}), but find results incompatible with
observations. Figure \ref{fig:both_low} shows the surface density and
temperature projections at the end of the simulation. While the height of the
bubbles is comparable to that found in the ``Base'' simulation, the width is
too narrow. Also the gap between the Galactic centre and the lower edge of the
bubbles is too large, and the morphology of the whole region appears too
non-uniform compared with the observations by \cite{Su2010ApJ}.

Increasing both $t_{\rm q}$ and $f_{\rm g}$  simultaneously by a factor
  larger than a few is also not a viable option. Morphologically, had the CMZ
been able to withstand onslaught from \sgra\ feedback for longer, the dynamics
of the gas may have produced reasonably shaped hot bubbles. But we are already
using a CMZ mass that  corresponds to the upper limits derived from
  observations \citep{Morris1996ARA&A}, and even with $t_{\rm q} = 1$~Myr the
CMZ is significantly affected in the ``Base'' simulation. It appears to us
that increasing $t_{\rm q}$ further may do too large a damage to the CMZ,
i.e., drive most of it to much larger distances from the Galactic Centre than
it currently observed. Therefore we feel that $t_{\rm q}$ longer than $\sim
1$~Myr are not very likely, although there remains a possibility that the
current CMZ is only a remnant of an even more massive gas disc that preceeded
the \sgra\ quasar outburst.

\subsection{Dependence on the CMZ properties} \label{sec:varhr}

Having checked the effects of varying the halo and quasar parameters, we now
vary the properties of the CMZ. In simulation `Mc-low', we reduce the CMZ mass
by a factor of 10, making it smaller than the current observational estimates
\citep[$M_{\rm cmz} \sim 3-5 \times 10^7
  \msun$;][]{Dahmen1998A&A,PiercePrice2000ApJ}.

The results are shown in Fig.  \ref{fig:vartests2}, left. Comparing the figure
with the right panel of Figure \ref{fig:Base_evol}, we note that the shape of
the bubbles and the shocked region is closer to spherical in the case of a
less massive CMZ. Although the lower CMZ mass still yields a weight $\sim5$
times larger than what the quasar outburst should be able to lift, the
vertical density stratification in a homogeneous disc makes a large part of
the CMZ diffuse enough to be blown away. The rest of the CMZ is shock-heated
to higher temperatures than in the `Base' simulation and begins to expand
vertically, providing more material that can be removed by the wind.  As a
result, most of the CMZ is blown away by $\sim 0.7$~Myr in simulation
'Mc-low', and the outflow proceeds quasi-spherically for the last $0.3$~Myr of
the quasar activity. Although almost all of this material eventually accretes
back onto the reforming CMZ disc after the quasar has switched off, the
effects on the diffuse ``halo'' gas are profound. The bubble height is
$\sim40\%$ lower, their width is $\sim20\%$ smaller and the bubbles are
significantly detached from the Galactic centre. This simulation does not
appear to match the morphology of the {\it Fermi} bubbles as well as the
``Base'' simulation does.

We next check the influence of the CMZ geometry by reducing its scale height
to yield the geometrical aspect ratio of $H/R = 0.125$, with a corresponding
decrease in the temperature floor of the simulation ('HR-low',
Fig. \ref{fig:vartests2}, middle). This value is smaller than the CMZ aspect
ratio favored by the current observations \citep[$H/R \sim
  0.15$;][]{PiercePrice2000ApJ,Jones2011MNRAS}, so that we effectively cover
all the reasonable parameter space in terms of $H/R$ for the CMZ. 

We find no qualitative difference between simulations `HR-low' and
'Base'. This suggests that even if the exact geometry of the CMZ plays a role
in determining the bubble shape, the magnitude of this effect is quite
limited. In all simulations, the bubbles are collimated much more strongly
than pure shielding by the CMZ would suggest (the opening angle of the bubbles
is $\Omega_{\rm b} \sim 0.4 \times 4\pi$, while the solid angle not obscured
by the CMZ in the `Base' simulation is $\Omega_{\rm cmz} \sim 0.8 \times
4\pi$), therefore this lack of difference is not particularly surprising.

\subsection{Effect of the heating-cooling prescription}

Simulation `Cool' is identical to the `Base' simulation, except that we use a
physically motivated optically thin quasar heating-cooling prescription
\citep{Sazonov2005MNRAS} instead of the standard
  \citet{Sutherland1993ApJS} one, as in the other simulations in this paper.
The prescription is based on a fit to the radiative heating and cooling rates
of the gas illuminated by a typical quasar radiation field. The radiative
processes include photoionization heating, Compton and inverse-Compton
processes, bremsstrahlung and line cooling, and assume an optically thin
medium, which is well justified for the low column densities that we find in
this paper, $\Sigma \lesssim 10^{-4}$~g cm$^{-2}$. The only region where gas
may become optically thick to X-rays from \sgra\ is the midplane of the CMZ
disc. However, we find that the higher density gas is able to cool efficiently
anyway, thus staying close to the imposed temperature floor of $T=$ a few
  times $10^4$~K in any event.

The overall effect of changing the heating-cooling prescription on the large
scale gas distribution is small (Fig. \ref{fig:vartests2}, right). The bubble
interior is $\sim50\%$ hotter ($T_{\rm bub} \sim 3\times10^8$~K as opposed to
$2\times10^8$~K in the `Base' simulation), most likely due to less efficient
cooling in the adopted prescription at $t > t_{\rm q}$. The temperature of the
outer shell is practically the same. Morphologically, this difference results
in bubbles that have $\sim10\%$ lower height and practically the same width,
although the KH-unstable ripples are missing. The bubbles are also somewhat
less collimated, as the material close to the Galactic plane is able to cool
down more efficiently and collapse to higher densities. In general, however,
the bubbles look rather similar.  This shows that our results are somewhat
insensitive to the details of gas heating and cooling.  This finding is also
consistent with the analytical predictions of \citet{King2010MNRASa} and
\citet{King2011MNRAS}, where by construction an energy-driven outflow occurs
when cooling of the shocked wind and the  shocked ISM becomes
inefficient; we comment on this result further in Section
\ref{sec:energy}. The effect of the X-ray heating from \sgra\ on the CMZ
material is even smaller than that on the diffuse ambient gas, as explained
above.

We point out that this insensitivity to quasar photo-ionisation heating and
radiative cooling is to be expected due to the low density and a rather short
duration of \sgra\ outburst compared with typical cosmological conditions
\citep{Sazonov2005MNRAS}. At higher densities we would expect the structure of
the ambient gas to be significantly dependent on the details of the
  cooling function employed in the simulations.

\section{Discussion} \label{sec:discuss}

\subsection{Summary of simulation results}

In general, our numerical simulations appear to confirm the suggestion of
\cite{Zubovas2011MNRAS} that an Eddington-limited outburst of \sgra\ outflow
is a promising way of explaining the morphology of the observed {\it Fermi}
bubbles. Starting with the model for AGN feedback developed to explain the
$M_{\rm bh}-\sigma$ relation for classical bulges and elliptical galaxies
\citep{King2003ApJ, King2005ApJ}, the ``typical'' initial conditions for
numerical simulations of AGN feedback \citep{Nayakshin2010MNRAS} needed to be
ammended only to account for (a) the lower present day gas content of the
Milky Way (i.e., smaller ``gas fraction'' $f_{\rm g}$); (b) the presence of a
massive gas disc, the CMZ, in the plane of the Galaxy in the central $200$~pc;
(c) a finite duration of \sgra\ outburst, which is a free parameter of our
model, and is small compared with a dynamical time in a host galaxy bulge.

Given this setup, we confirm that a spherically symmetric outflow from
\sgra\ is collimated by a geometrically thin CMZ disc in directions
perpendicular to the Galactic plane. The outflow then produces two teardrop
shaped cavities that have sizes similar to the observed $\gamma$-ray emission
features. We varied the free parameters of our model - the quasar outburst
duration $t_{\rm q}$ and the gas fraction $f_{\rm g}$ - to constrain their
values to $t_{\rm q} \approx 1$~Myr and $f_{\rm g} \approx 10^{-3}$. The
former is plausible and has interesting implications for \sgra\ feeding
(cf. \S \ref{sec:feeding} below). The latter is poorly constrained
observationally but consistent with estimates by \citet{McKee1990ASPC} and
\citet{Sofue2011arXiv}.

The opening angles of the bubbles as seen from \sgra\ are $\Omega_{\rm b} \sim
0.4 \times 4\pi$. This is significantly smaller than the solid angle not
obscured by the CMZ in the `Base' simulation ($\Omega_{\rm cmz} \sim 0.8
\times 4\pi$), showing that CMZ casts a larger ``feedback shadow'' than could
be expected based on its geometrical aspect ratio alone. This is driven by the
following two effects. First of all, the CMZ not only hinders the outflow
propagation directly through it, but also ``reflects'' part of the
outflow. The thermally driven outflow of hot gas ablated from the CMZ surfaces
away from the Galactic plane redirects the gas flowlines towards vertical
directions. Secondly, the bubbles rise due to buoyancy, and cooler material
streams to fill the void along the Galactic plane, further reducing the
opening angle of the bubbles.

In addition, although we were not originally interested in the evolution of
the CMZ due to \sgra\ feedback -- the role of the CMZ in our simulations was
to stop and redirect the quasar wind only -- we found several CMZ-related
results interesting from an observational point of view: (a) re-shaping of the
CMZ into a ring-like structure, perhaps explaining Herschel observations that
CMZ is a dusty ring rather than a disc; (b) an induced star formation
mode in the CMZ which is trigerred by \sgra\ feedback. These points are
further discussed in \S \ref{sec:cmzshape}.

We now make a detailed comparison of our results to observations.

\subsection{Gas mass within the bubbles} \label{sec:mass}

In our simulations, two cavities corresponding to the observed {\it Fermi}
bubbles are filled with hot and diffuse gas. The observationally estimated
mass of gas within the bubbles \citep[$M_{\rm bub} \sim 10^8 \;
  \msun$;][]{Su2010ApJ} is much greater than the mass of the shocked wind in
our model. The latter, from analytical arguments in Section \ref{sec:model},
is $M_{\rm w} \simeq \dot{M}_{\rm Edd}t_{\rm q} \sim 8 \times 10^4 \; \msun$
for $t_{\rm q} = 1$~Myr. The mass contained inside the cavities is $M_{\rm
  bub,sim} \sim 6 \times 10^5 \msun$ for the `Base' simulation. The density of
gas within the bubbles ($n_{\rm bub} \sim 3\times10^{-5}$~cm$^{-3}$) is also
much lower than the one typically adopted in spectral modelling of the bubble
emission \citep[$n \sim 10^{-2}$,][]{Su2010ApJ, Crocker2011PhRvL}.

We believe that our simulations underpredict the gas density inside the
bubbles which would otherwise be obtained in a more sophisticated
simulation. We use a one-phase model for the ambient medium, whereas
observations of gas in the Milky Way show that it is multi-phase
\citep{Dame2001ApJ}. We would thus expect some cold and warm gas to be present
even at large heights above the disc before \sgra\ ``turns on''. Much like in
supernova shocks expanding into the ambient medium, molecular clouds embedded
in the ambient gas are expected to be overtaken by the quasar wind and later
evaporate inside the bubbles, increasing the hot gas density there at late
times \citep[e.g.][]{McKeeCowie75}. In addition to that, having strong density
inhomogeneities in the ambient medium should provoke strong Rayleigh-Taylor
instabilities \citep{King2010MNRASb} during the quasar outburst event, which
we do not model here. These may allow formation of dense filaments
resilient to \sgra\ feedback that are left behind the shock just like the
molecular clouds discussed above.  We hope to explore these ideas with
improved simulations in the future.

\subsection{Bubble energy content} \label{sec:energy}

\begin{figure}
  \centering
  \includegraphics[width=0.45\textwidth]{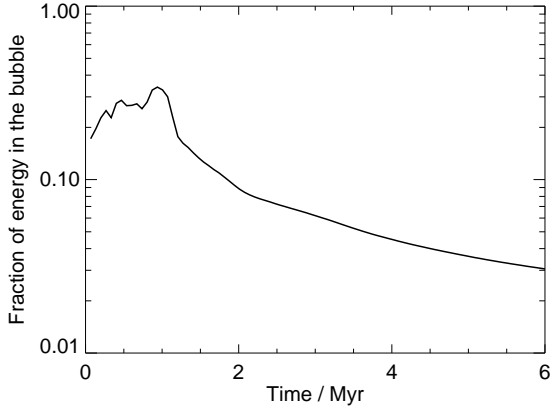}
  \caption{Time evolution of the fraction of energy input into the system by
    the quasar wind that is retained inside the bubbles. The rest of the
    energy is contained in the shell of shocked expanding ambient gas.}
  \label{fig:buben}
\end{figure}

The total energy input by the quasar into the system is 
\begin{equation}
E_{\rm in} \simeq \frac{\eta}{2}\; L_{\rm Edd}\; t_{\rm q} \simeq 8 \times
10^{56} \; t_6 \; {\rm erg}, 
\end{equation}
where $t_6$ is quasar activity time in Myr and $\eta / 2 = 0.05$ is the
coupling efficiency of the energy-driven wind (eq. \ref{eq:dote}). As
mentioned in \S \ref{sec:model}, $E_{\rm in}$ is significantly larger than the
observational estimate of the total gas energy content of the bubble: $E_{\rm
  bub, obs} \sim 10^{54-55}$~erg \citep{Su2010ApJ}. However, in our
simulations, the injected energy $E_{\rm in}$ is split between the bubble and
the surrounding shell. Even when the quasar is active, analytical calculations
show that only $1/3$ of the input energy remains inside the bubble
\citep{Zubovas2012ApJ}, and this value should decrease after the quasar has
switched off, as the bubble cools due to expansion.

To quantify this further, we computed the fraction of energy injected into the
ambient gas by \sgra\ and retained by the bubble in the `Base' simulation as a
function of time (other simulations show similar results). To this end, we
first compute the sum of kinetic and thermal energy for all the gas particles
in the simulation, $E_{\rm tot}(t)$.  We exclude in this calculation the
gravitational potential energy of gas because for most SPH particles in the
calculation changes in the latter are small compared with the changes in the
kinetic and thermal energy. We then define the energy content of the gas
within the bubble, $E_{\rm bub}(t)$, as the sum of kinetic and thermal energy
of the gas inside the bubble.  The fraction of energy retained in the bubble
is thus defined as
\begin{equation}
\epsilon_{\rm E,in}(t) \equiv \frac{\Delta E_{\rm bub}}{\Delta E_{\rm tot}} =
\frac{E_{\rm bub}(t) - E_{\rm bub}(0)}{E_{\rm tot}(t) - E_{\rm tot}(0)}\;.
\end{equation}

Figure \ref{fig:buben} presents the evolution of $\epsilon_{\rm E,in}$ with
time. While the quasar is active, the fraction of energy retained inside the
bubble grows slightly and fluctuates around the analytically derived value
of $1/3$. Once the quasar switches off, the expanding bubble cools and
transfers most of its energy to the surrounding shell, so that by $t = 6$~Myr,
only $\sim 3\%$ of the total energy input is retained inside the bubble. Hence
the total energy content of the bubbles by the end of the simulation is
\begin{equation}
E_{\rm bub} \sim 0.03 E_{\rm in} \simeq 2.5 \times 10^{55} \; t_6 \; {\rm
  erg},
\label{ebub_base}
\end{equation}
a value similar to but still a little larger than the observed one.

A further decrease in the bubble energy content may be caused by
  radiative cooling. To check this, we consider the evolution of $E_{\rm
    tot}(t)$ and find that after $t=t_{\rm q}$ it is conserved to within a few
  $\%$. Adding a reasonably parametrized (using $r_{\rm s} = 1$~kpc as a scale
  radius) value for the gas potential energy does not change the result
  either. Therefore the importance of cooling on the total energy content is
  minimal and does not affect our previous considerations.

\subsection{Expected radiation from the bubble} \label{sec:radiation}

The main emission components of interest are the $\gamma$-ray emission from the
the lobes \citep{Su2010ApJ}, a tentative microwave feature coincident with the
bubbles identified in the WMAP all-sky maps
\citep{Finkbeiner2004ApJ,Su2010ApJ} and an X-shaped feature visible in the
X-rays closer to the Galactic plane, coinciding with the edges of the bubbles,
known as the ``ROSAT limbs'' \citep{Snowden1997ApJ}. We discuss each of the
three components in turn.

While we do not model cosmic ray (CR) particles in our paper, we note that
astrophysical shocks are known to accelerate electrons, protons and other
particles to CR energies \citep{Blandford1987PhR} in a variety of environemnts
and put as much as $\sim 10$\% of the blast wave energy into the high energy
particle component in the case of supernova shocks. We certainly do not see an
obvious reason why shocks driven by an even faster outflow from a quasar would
be less efficient in producing CRs than supernovae.

Two competing explanations for the origin of the $\gamma$-ray emission from
the {\it Fermi} bubbles were suggested in the literature to
date. \citet{Crocker2011PhRvL} suggested that the emission is powered by CR
protons through {\it pp} collisions with the plasma in the bubbles. In this
scenario, $\sim 10^{39}$erg~s$^{-1}$ of energy is injected into the bubbles in
the form of CRs for $\sim 10^{10}$ yrs to achieve saturation in the
  system \citep[cf. also][]{Crocker12a}. This yields $\sim 3\times 10^{58}$
erg in CR enegy alone. This is some 3 orders of magnitude larger than the
thermal energy retained within the bubbles in our ``Base'' simulation
(equation \ref{ebub_base}). Therefore we conclude that our model is very
unlikely to produce a bright enough $\gamma$-ray emission if emitting
particles are hadrons.

On the other hand, \citet{Mertsch2011PhRvL} explore the possibility that
electrons accelerated by shocks in turbulent plasma inside the bubbles can
reproduce the spectral features, as well as the constant surface brightness
profile. They find that this acceleration process can continuously replenish
the energetic electron population, overcoming the problem of rapid electron
cooling via the inverse-Compton process. The cooling time due to both IC and
synchrotron losses is less than $5$~Myr for $100$~GeV electrons at $z=5$~kpc
above the Galactic plane \citep[Fig. 28 and Section 7.1
  in][]{Su2010ApJ}. Since the current $\gamma$-ray luminosity of the bubbles
is $\sim 4\times 10^{37}$ erg~s$^{-1}$, and the age of the bubbles is $6$ Myrs
in our model, we conclude that at least $\sim 10^{52}$ erg of CR electrons is
required to explain the {\it Fermi} Bubbles, and somewhat more if we account for
continuous replenishment from plasma turbulence. This would be a very small
fraction of the bubble energy content in our model, and on that basis at least
is possible.

 The origin of ``continuous'' shock driving inside the bubbles can be
  naturally explained by our model. Since the {\it Fermi} bubbles are still
  expanding at the end of the simulations, we expect that the whole volume of
the cavities inflated by the quasar outburst is still filled with shocks. As
noted in \S \ref{sec:mass}, a higher resolution modelling that includes a
multi-phase description for the ambient pre-shock gas is likely to result in
strong Rayleigh-Taylor instabilities of the wind shock front during the quasar
outburst event \citep{King2010MNRASb}. This should let cold gas filaments fall
into the bubble's interior due to gravity, leading to high Mach number
shocks. Protons and electrons could thus be continuosly accelerated on these
shocks now, as suggested by \citet{Mertsch2011PhRvL}.

The observed microwave haze spatially coincides with the {\it Fermi} bubbles,
although its intensity decreases with height above the Galactic plane,
especially in the southern bubble \citep[see the bottom right panel of Figure
  18 in][]{Su2010ApJ}. The luminosity of synchrotron emission from cooling
electrons decreases with height above the plane possibly due to an expected
decrease in the magnetic field strength. We believe that same electrons
responsible for the $\gamma$-ray emission may be responsible for the WMAP
haze.

The ROSAT-observed X-ray background in the region of the {\it Fermi} bubbles is
composed of X-shaped ridges coincident with the lower limbs of the bubbles,
most clearly visible in the hard ($1.5$~keV) band of the instrument, and a
cavity in soft-band X-ray emission coincident with the bulk of the bubble area
\citep{Snowden1997ApJ, Su2010ApJ}. The spatial coincidence of the features
suggests a common origin.

In our simulations, gas inside the bubbles has temperatures as high as
$\sim10-20$~keV (see Figure \ref{fig:Base_prop}, right) and so most of its
thermal emission is harder than what ROSAT would detect. This could then
explain the deficit of soft X-ray emission from within the bubbles.  The
shocked ISM (see Section \ref{sec:ismshock} below) is heated to temperatures
of a few times $10^7$~K, (although that is expected to scale as $\propto
\sigma^2$, so may vary somewhat if the model for the potential in the inner 10
kpc of the Galaxy is varied). The shocked shells around {\it Fermi} bubbles
should then be most visible in the harder ROSAT bands. X-ray emissivity scales
as $\propto n^2$, where $n$ is gas particle density. The latter tends to be
larger close to the Galactic plane due to superposition of the shells around
the two bubbles (Figure \ref{fig:Base_evol}, right) and vertical
stratification.  Therefore we may expect the lower limbs of the shells to be
brighter in X-rays, as observed.

\subsection{Bubble edges and the outer ISM shock} \label{sec:ismshock}

The expanding SMBH wind drives a strong forward shock into the ISM, which is
expected to be adiabatic and move with $v_{\rm ISM} \sim 4/3 v_{\rm e}$
\citep{Zubovas2012ApJ}. Since the shell thickness is initially zero, its
  vertical extent should also be $r_{\rm ISM} \sim 4/3 r_{\rm b}$, i.e. the
  thickness of the snowplough shell should be $1/3$rd of the bubble height
\citep{Zubovas2012ApJ}. We note that in the simulations, it is somewhat
greater: $d_{\rm sh} / r_{\rm b} \sim 0.4$. This result is not due to poor
resolution.  At late times, the thickness of the shell is $\sim10$ times
greater than the typical SPH particle smoothing length inside it ($h_{\rm
  SPH,sh} \sim 500$~pc), therefore we believe the large-scale morphology of
the region is well resolved.

The most likely reason for the discrepancy is the fact that the analytical
model assumes a continuous quasar outflow driving the shell, whereas in our
simulations the quasar turns off at $t= 1$~Myr. As a result the shell is less
compressed in the radial direction.

In addition, our one-phase treatment of the ambient diffuse medium (cf. \S
\ref{sec:mass}) can under-estimate the radiative cooling in the
shell. Therefore, a more realistic multi-phase simulation could be expected to
have thinner shell enveloping the {\it Fermi} bubbles.

Simulations also show transition regions at the edges of the bubbles ( e.g. at
$r_{\rm cyl} = 4-5$~kpc in Figure \ref{fig:Base_prop}, right). These are
$\lesssim1$~kpc thick zones where both temperature and density change from the
values appropriate for bubble interior to those of the swept-up ISM. The
observed bubble edges are $5-10^{\rm o}$ wide, corresponding to a few hundred
parsecs, in good agreement with simulation results. However, we must note that
the SPH particle smoothing length in these regions is comparable to the region
thickness, therefore numerical effects probably dominate the result.

\subsection{Has \sgra\ feedback affected the CMZ?} \label{sec:cmzshape}

The primary goal of our paper is investigating whether \sgra\ feedback is a
reasonable model for the morphology of the {\it Fermi} bubbles. In doing so we
purposely introduced a very simple model for the CMZ -- a $\Sigma(R) \propto
R^{-1}$ circular disc in the plane of the Galaxy with mass and size
consistent with observations \citep{Morris1996ARA&A}. This simple model
nevertheless resulted in an interesting transformation and exitation of the
CMZ that is worthy of further discussion.

\subsubsection{The Herschel ring: a feedback-compressed disc?}\label{sec:compress}

In particular, we find that, under the strong coercion from \sgra\ feedback,
the initial disky configuration of the CMZ attains a morphology more
reminiscent of a dense ring (see Section \ref{sec:Base_centre} and Figure
\ref{fig:Base_centre}). We believe this is a general result of an AGN feedback
acting on a disk since the physics behind this ``anti-diffusion'' evolution of
the disk is simple and thus robust.

When the inner regions of the disc are blasted with an outward-directed
feedback from the centre, little of the material outflows to infinity; most
actually falls back onto the disc at large radii, mixes with gas there and
thus induces an inward-directed radial flow at those radii. AGN feedback thus
serves as an external agent that forces shock mixing of gas initially located
at different radii and carrying different specific angular momenta. An
initially broad distribution of angular momentum (a disc) becomes more narrow
(a ring).

The slow viscous evolution of the dense ring that is formed by feedback
suggests that such a ring should be still present in the GC today.  In fact,
recent {\it Herschel} observations of the GC region have revealed a ring-like
structure on the scale of $\sim100$~pc \citep{Molinari2011ApJ}, qualitatively
similar to what we see in the simulations. The observed feature is elliptical
and offset from \sgra. In Section \ref{sec:offset} below we discuss how this
strongly non-circular and offset structure could be formed in our model.

The fact that the CMZ is only perturbed but not dispersed to infinity is
consistent with analytical predictions in \S \ref{sec:waist}. We note also
that it may seem paradoxical how this very massive gas disc gets evacuated in
the inner region and yet \sgra\ continues to accrete from presumably a much
smaller disc at the assumed Eddington rate for $\sim 1$~Myr as in our ``Base''
simulation. However, the paradox is easily solved by realising that weight per
unit mass of the gas -- i.e., acceleration due to gravity -- scales as $\propto
M(R)/R^2$, where $M(R)$ is the total mass enclosed within radius $R$. The disc
that fed \sgra\ may have had a radius of order $R\sim 0.1$ pc or even less
\citep{Nayakshin2005A&A,AlexanderEtal12a}. The weight of the {\em accretion
  disc} of mass $M_{\rm d} = 10^5 M_5 \msun$ is then
\begin{equation}\label{wdisc}
W_{\rm ac} = \frac{G\mbh M_{\rm d}}{R} = 1.2\times 10^{36} M_5\;\hbox{dyn}\;,
\end{equation}
which is some $\sim 20$ times larger than that estimated for the CMZ (equation
\ref{wcmz}). It is thus perfectly reasonable that a smaller-scale disk would
be too tightly bound to the SMBH to be expelled or even significantly
perturbed by feedback, unlike gas at larger radii.  Nayakshin, Power \& King
(2012, ApJ submitted) show in their Appendix that this is a general point.

\subsubsection{Non-circular orbit of the Herschel ring}\label{sec:offset}

\cite{Molinari2011ApJ} find that the Herschel ring is offset from \sgra\ and
thus the centre of the Milky Way by about 50 pc, which is comparable with the
size of the ring itself (see their figure 5). Any offset, and especially such
a large one, is not naturally expected if the ring is a long-lived structure.
While our model shows no offset, we note that there may be a natural way to
produce that, although further numerical simulations are needed to confirm
these ideas.

In particular, we have assumed here that the feedback from \sgra\ is exactly
spherically symmetric. This is likely to be an over-simplification. Even if
wind outflow is quasi-spherical on small scales (e.g., hundreds of
Schwarzchild radii), on the somewhat larger scales of stellar discs we know
that there must be a an accretion disc which cannot be readily expelled by
\sgra\ feedback as discussed above. We also know that the young stellar discs
orbiting \sgra\ are not simple planar structures, with significant warps
needed to explain the observed stellar kinematics \citep{Bartko2009ApJ}. Large
warps also naturally occur in numerical simulations of star forming gas flows
\citep{Hobbs2009MNRAS}. Therefore, due to shadowing of \sgra\ feedback by the
$\sim 0.1$ pc gas flow (which may be strongly non-circular), one side of the
CMZ may have experienced a different amount of feedback compared with the
other side, causing a strong non-axysimmetric perturbation to the CMZ disc and
perhaps leading to an offset ring reminiscent of the observed Herschel
ring. We note that the asymmetry in the feedback should not be too large over
the whole $4\pi$ solid angle, however, so as to not result in too asymmetric
{\it Fermi} Bubbles.

Additionally, natural non-axisymmetric density variations in the pre-feedback
CMZ may result in outflow within the CMZ proceeding at different velocities in
different directions, producing the offset.

\subsubsection{Induced star formation in the CMZ?}\label{sec:induced}

A particularly interesting outcome of the CMZ compression by the outflow is
the formation of a dense clump of gas at $t\sim3.5$~Myr (see Section
\ref{sec:Base_centre}). The large mass, $\sim10^7 \msun$, of the clump and its
position at $R\sim100$~pc from the Galactic centre suggest that the observed
young Galactic centre stellar clusters
\citep{Serabyn1998Natur,Figer1999ApJa,Figer1999ApJb} or some of the largest
molecular clouds \citep[e.g. Sgr B2,][]{Scoville1975ApJ, Reid2009ApJ} may have
formed this way. Many smaller clumps in the dense ring may be susceptible to
star formation; recent IR observations \citep{Yusef-Zadeh2009ApJ,Immer2012A&A}
reveal that the star formation rate in the CMZ during the past $\sim1$~Myr was
$\sim 0.08 \msun$~yr$^{-1}$. At this rate, all of the CMZ gas would be turned
into stars in $\sim 6 \times 10^8$~yr; even if the CMZ was supplied by
molecular gas from further out, the whole molecular gas content of the Milky
Way would have been turned into stars in $\sim 6$~Gyr. This suggests that the
current star formation rate in the CMZ is higher than the long-term average,
further implying a rather recent perturbation, consistent with the results of
our simulations. Furthermore, combined {\it Chandra} and {\it HST}
observations have revealed a population of isolated massive (O and B giant,
but also WR) stars in the CMZ \citep[e.g.][]{Mauerhan2010ApJ}. These stars are
not associated with any of the known clusters and may have formed in small
associations. Our simulations suggest that while these stars are isolated now,
they may have formed coherently in time if not in space due to a single very
powerful perturbation produced by \sgra\ quasar outburst that turned the disc
into a star-forming ring.

Finally, we note that the strong perturbations to the initially circular
orbits of gas in the CMZ by \sgra\ feedback could form gas clumps on eccentric
orbit. Indeed, the orbit of the clump found in the ``Base'' simulations is
mildly eccentric, $e \sim 0.2$. That was obtained in a perfectly azimuthally
symmetric simulation; any deviation from this assumption would have likely
resulted in an even more eccentric clump. Such eccentric clump formation
mechanism may be relevant to the origin of the Arches cluster which has a
rather non-circular orbit \citep{StolteEtal08}.

\subsection{Implications for AGN feeding models}\label{sec:feeding}

\sgra\ is the closest SMBH, and while its dimness is a familiar (and
important) tale that stimulated development of non-radiative models of
accretion for low density gas flows near SMBHs
\citep[e.g.,][]{NarayanEtal95,BB99}, little has been known about the past
  of \sgra\ as an accretion-powered SMBH. This is mainly due to potential
difficulties of discovering signs of past \sgra\ activity. Light-echos due to
X-ray fluorescence of \sgra\ radiation on molecular clouds is a powerful
technique \citep{Sunyaev1998MNRAS,Revnivtsev2004A&A,Terrier2010ApJ} but can
only be used to constrain \sgra\ activity up to $\sim 10^4$ yrs at best due to
the finite size of the Galaxy and the $\sim 1/R^2$ fading of signals from
clouds at large distance $R$ from the Galactic Centre.

However, shocks induced by an outflow from \sgra\ may continue to be visible
for about a dynamical time of the Galaxy, i.e., $R_{\rm G}/\sigma\sim 50$
Myrs, where we set Galaxy ``radius'' to be $\sim 5$~kpc and velocity
dispersion $\sigma = 100$ km~s$^{-1}$ for illustrative purposes. Therefore,
the {\it Fermi} lobes detected by \cite{Su2010ApJ} may be such a
``shock-echo'' of the past \sgra\ activity. Following \cite{Zubovas2011MNRAS}
we required the activity episode of \sgra\ to coincide with the well known
star formation event in the Galactic Centre $\sim 6$ million years ago
\citep{Paumard2006ApJ}, and found that the model has a number of attractive
observational consequences which lend some support to this picture.

If \sgra\ indeed did have an Eddington-limited outburst for as long as $\sim
1$~Myr, it must have accreted $\sim 10^5 \msun$ of gas. This is an order of
magnitude more than the mass in the stellar discs
\citep{Paumard2006ApJ,Nayakshin2006MNRAS}. The fact that \sgra\ managed to
accrete $\sim 90$\% of the gas from the star-forming accretion disc is highly
significant for the general question of how SMBH are fed, and should be
explored further as a potential example of an AGN disc that avoided ``star
formation catasrophe'' in which gas is believed to be turned into stars too
rapidly to feed AGN \citep{Goodman2003MNRAS,Nayakshin2007MNRAS}.

\subsection{Uncertainties and deficiencies of our work}\label{Sec:def}

As discussed in \S \ref{sec:radiation}, we model the ambient gas ``halo'' with
a single phase medium, whereas the ISM is expected to be multi-phase
\citep{McKee1990ASPC}. The lower density medium, essentially unresolved in our
simulations, is probably quite important for the dynamics of the fast outflow
from \sgra\ interacting with the gas in the bulge of the Galaxy (see also \S
\ref{sec:mass}). Therefore some of our conclusions (e.g., the likely quasar
phase duration $t_{\rm q}$, the geometrical thickness of the shocked shell,
etc.)  may somewhat depend on the treatment of the ambient medium, and future
work is needed to quantify this issue.

Also, a more realistic model for the geometrical arrangement of the diffuse
``halo'' gas that the outflow interacts with is desirable as this probably
affects the eventual shape of the bubbles. In our simulations, we consider a
spherical halo mass distribution, whereas in reality, we expect stratification
on a large scale, with higher gas density in the Galactic plane than
perpendicular to it, and higher density in the bulge than outside. We can make
a very simple estimate of the magnitude of this effect by considering the
dependence of outflow stalling radius on the gas fraction, which we take here
to be the mean gas fraction along a direction of expansion. The analysis in
Paper I and Section \ref{sec:model} shows that $R_{\rm stall} \propto v_{\rm
  e}^2 \propto f_{\rm g}^{2/3}$. As the bubble expands in an approximately
self-similar fashion (see Section \ref{sec:Base}) after the quasar switches
off, we can estimate that in any given direction $R_{\rm bub} \propto f_{\rm
  g,eff.}^{2/3}$. Therefore a factor of three difference in gas fraction
results in approximately a factor of two difference in bubble radius. We hope
that future observations of diffuse gas in the bulge of the Milky Way will
help to constrain this part of our model.

Our results are, as expected, sensitive to the total mass of the CMZ,
producing bubble morphology inconsistent with observations if it is reduced to
$10^7 \msun$ or less. This low value of the CMZ mass is however
unrealistically small, smaller than the current observational constraints of
$3-5 \times 10^7 \msun$ \citep{Dahmen1998A&A,PiercePrice2000ApJ}.  A
potentially important caveat is that we model the CMZ as a smooth disc, rather
than as a clumpy distribution of gas \citep{Molinari2011ApJ,
  Morris1996ARA&A}. In principle, a real outflow may stream past the clumps
and so the collimation effect of the CMZ would be smaller than what we
find. However, observations also show that the column density of gas toward
the Galactic centre does not vary strongly with viewing direction
\citep{Goto2008ApJ}, so the CMZ should provide a strong covering effect to any
outflow coming from \sgra. In addition, we find that the CMZ aspect ratio has
a negligible effect on the bubble properties, i.e. there is a range of CMZ
column densities for which the outflow is collimated efficiently. Therefore,
we believe that the uncertainty in our results due to the shape of the CMZ is
relatively small.

\section{Conclusions} \label{sec:concl}

We have presented numerical simulations of a wide angle outflow from \sgra,
temporally coincident with the star formation event $\sim6$~Myr ago, and
collimated into directions perpendicular to the plane of the Galaxy by the
presence of a massive disc of gas, the Central Molecular Zone. Our main
results are:

\begin{itemize}
\item This model is a plausible way to inflate the $\gamma$-ray emitting lobes
  recently observed by the {\it Fermi}-LAT, provided that \sgra's outburst
  duration is $\simlt 1$~Myr. 
\item The energetics of the model is consistent with the observed $\gamma$-ray
  emission if radiating particles are electrons rather than hadrons.
\item \sgra\ feedback could have reshaped the CMZ from a disc-like
  configuration into a ring-like one reminiscent of the observed structure
  \citep{Molinari2011ApJ}. We speculate that somewhat assymetric feedback
  could produce an observed offset of the ring and eccentric orbits for young
  star clusters such as the Arches star cluster.
\item Furthermore, CMZ compression by the feedback outflow may explain the
  formation of dense GMCs, the young star clusters, the population of isolated
  massive stars, and also result in the present-day high star formation rate
  in the Galactic Centre.

\end{itemize}

An important side result of this paper is that the same feedback model appears
to work for quasars as their SMBHs establish the $M-\sigma$ relation and clear
the host galaxies of gas
\citep[e.g.][]{King2003ApJ,King2005ApJ,Nayakshin2010MNRAS}, as well as for a
short burst of activity in \sgra, a SMBH that is somewhat below the
$M-\sigma$ relation in a quiescent galaxy. This finding suggests that there is
nothing fundamentally different between feedback from SMBHs at gas-rich epochs
($z \gtrsim 2$) and that from local galactic nuclei, except for the lower
amount of fuel they receive.

More detailed future treatment of the feedback process, with improvements in
both the physics of the simulations and more realistic observationally
constrained initial conditions, may provide interesting and unique constraints
on cosmological models of AGN feedback.

\section*{Aknowledgments}

This research used the ALICE High Performance Computing Facility at the
University of Leicester.  Some resources on ALICE form part of the DiRAC
Facility jointly funded by STFC and the Large Facilities Capital Fund of BIS.

KZ is supported by an STFC studentship. Theoretical astrophysics research in
Leicester is supported by an STFC Rolling Grant.

\end{document}